\documentclass[aps,pra,10pt,groupedaddress,notitlepage,showpacs,floatfix,longbibliography,superscriptaddress,twocolumn,nofootinbib]{revtex4-2}

\usepackage[subrefformat=parens,labelformat=parens]{subcaption}
\usepackage{ragged2e}
\usepackage{tikz}
\usetikzlibrary{fit, positioning} % fit is needed for the boxes, positioning often helps

\usepackage{mathtools}
\usepackage{braket}
\usetikzlibrary{arrows.meta}
\usetikzlibrary{patterns}
\usetikzlibrary{calc} 
\usetikzlibrary{decorations.markings}
\usetikzlibrary{decorations.pathreplacing,calc, positioning}

\usepackage{pgfplots}
\usepackage{amsmath}
\usepackage[utf8]{inputenc}
\usepackage{graphicx}

\usepackage{array}[=2016-10-06]\usepackage{dcolumn}

\usepackage{bm}
\usepackage{mathrsfs}
\usepackage{amsfonts,amssymb}
\usepackage{color}
\usepackage{verbatim}
\usepackage[dvipsnames]{xcolor}

\usepackage{algorithm}
\usepackage[noend]{algpseudocode}

\usepackage{eqparbox}
\usepackage{dsfont}
\usepackage{braket}
\usepackage{algorithm}
\usepackage{float}
\usepackage{algpseudocode}
\usepackage{multirow}
\usepackage[english]{babel}
\usepackage{url}
\usepackage{hyperref}
\usepackage{amsthm}  % optional, if you want theorem-like environments
\usepackage[normalem]{ulem}
\definecolor{darkblue}{rgb}{0,0.3,0.7}
\hypersetup{
    colorlinks=true,
    linkcolor=darkblue,
    filecolor=blue,
    citecolor=darkblue,
    urlcolor=darkblue,
}
\usepackage{natbib}
\usepackage{cancel}

\usepackage[capitalize]{cleveref}
\crefname{section}{Sec.}{Secs.}
\Crefname{section}{Section}{Sections}

\crefrangelabelformat{section}{#3#1#4--#5\crefstripprefix{#1}{#2}#6}
\crefrangelabelformat{figure}{#3#1#4--#5\crefstripprefix{#1}{#2}#6}
\crefrangelabelformat{equation}{(#3#1#4--#5\crefstripprefix{#1}{#2}#6)}

\crefmultiformat{equation}%
{\edef\crefstripprefixinfo{#1}Eqs.~(#2#1#3}%
{,#2\crefstripprefix{\crefstripprefixinfo}{#1}#3)}%
{,#2\crefstripprefix{\crefstripprefixinfo}{#1}#3}%
{,#2\crefstripprefix{\crefstripprefixinfo}{#1}#3)}
\definecolor{teal}{RGB}{26,157,150}

\definecolor{purple}{RGB}{106,13,173}

\definecolor{maroon}{RGB}{130,0,0}

\newcommand{\imag}{\mathrm{i}}

\newcommand{\cF}{\mathcal{F}}
\newcommand{\cH}{\mathcal{H}}
\newcommand{\cR}{\mathcal{R}}

\definecolor{color1bg}{HTML}{1f77b4}
\definecolor{color2bg}{HTML}{ff7f0e}
\definecolor{color3bg}{HTML}{2ca02c}
\definecolor{color4bg}{HTML}{d62728}
\definecolor{color5bg}{HTML}{9467bd}
\definecolor{color6bg}{HTML}{8c564b}
\definecolor{colorfill}{HTML}{fbf19a}

\pgfplotsset{
  /pgfplots/error bar legend/.style={
    legend image code/.code={
        \draw[sharp plot,mark=-,mark repeat=2,mark size = 4 pt, line width = 2pt, mark phase=1,color=#1]
        plot coordinates { (0.3cm, -0.15cm) (0.3cm,0cm) (0.3cm, 0.15cm) };%
        
}}}

\pgfplotsset{compat=1.18}
\usepgfplotslibrary{groupplots}
\usepackage[color,frame,line,arrow,matrix,tips]{xy}

\pgfmathdeclarefunction{complexreal}{1}{%
    \pgfmathtruncatemacro{\posplus}{findpos("+",#1)}%
    \pgfmathparse{substring(#1,2,\posplus-2)}%
}

\begin{document}

\title{Re-examining the Role of State Texture in Gate Identification and Fixed-Point Resource Theories}

\author{Alexander C. B. Greenwood}
\affiliation{
Dept of Electrical \& Computer Engineering, University of Toronto, Toronto, Ontario, Canada M5S 3G4
}
\email{alexander.greenwood@mail.utoronto.ca}

\author{Joseph~M. Lukens}
\affiliation{Elmore Family School of Electrical and Computer Engineering and Purdue Quantum Science and Engineering Institute, Purdue University, West Lafayette, Indiana 47907, USA}
\affiliation{Quantum Information Science Section, Oak Ridge National Laboratory, Oak Ridge, Tennessee 37831, USA}

\author{Li Qian}
\affiliation{
Dept of Electrical \& Computer Engineering, University of Toronto, Toronto, Ontario, Canada M5S 3G4
}

\author{Brian T. Kirby}
\affiliation{DEVCOM Army Research Laboratory, Adelphi, MD 20783 USA}
\affiliation{Tulane University, New Orleans, LA 70118 USA}

\date{\today}

\begin{abstract}
A protocol for identifying controlled-NOT (CNOT) gates versus single-qubit-only gates in universal quantum circuits using randomized input states was recently shown to be intimately connected to the quantum resource of state texture. Here we revisit this gate identification protocol and demonstrate that a more general fidelity-based formulation succeeds for nearly all laboratory bases.
We then examine a broader family of quantum resource theories, where a distinct resource theory can be defined for each choice of reference pure state, establishing core resource-theoretic requirements without the computational shortcut offered by the ``grand sum'' employed in the original formulation of state texture. By extending from single ``resourceless'' states to convex sets via a convex-roof construction, we recover single-qubit measures of known resource theories such as imaginarity and coherence. 
Finally, we introduce a family of ``fixed-point resource theories'' that includes fixed-point instances of the theories of state texture, genuine coherence, purity, and athermality.
For these fixed-point resource theories we show that, under free operations, the fidelity-based lower bound is weakly monotonic, while specific violations of strong monotonicity are found for the convex-roof logarithmic measure.
\end{abstract}

\maketitle

\section{Introduction}

Quantum resource theories provide a framework for quantifying the operational value of physical properties central to quantum information processing \cite{chitambar2019quantum,coecke2016mathematical,costa2020information}. 
The general structure of resource theories consists of the definition of ``free'' resource states and operations (i.e. those that provide no operational value for a particular task), and ``resourceful'' states and operations. 
Free operations are resource-nonincreasing, and free states comprise those containing zero resource. 
In contrast, resourceful states and operations are those which are not free and, usually, singled out for their utility in a task which could not be performed with free states and operations alone.
Examples of quantum phenomena that have been cast as resource theories include entanglement \cite{horodecki2009quantum}, coherence \cite{streltsov2017colloquium, baumgratz2014quantifying, de2016genuine, marvian2016quantify, napoli2016robustness}, imaginarity \cite{wu2021resource}, and magic states \cite{veitch2014resource}, among several others \cite{chitambar2019quantum}. 

\begin{figure}[t!]
\centering

% Def Hilbert space
\tikzset{
  blob/.style={draw=black,fill=black!5,line width=0.9pt},
  face/.style={line width=1pt,black},
  iso/.style={very thin,black!60},
  point/.style={circle,fill=black,inner sep=1.8pt},
  zeropt/.style={circle,draw=black,fill=white,inner sep=1.2pt},
}
\def\BlobPath{%
  ( {3.10+5.2*cos(200)}, {0.40+5.2*sin(200)} )
   arc (200:160:5.2)
  .. controls (-1.00,2.20) and ( 2.20,2.20) .. ( 2.90,1.05)
  -- (3.10,0.40)
  -- (2.20,-1.40)
  -- (-0.40,-1.80) 
-- ( {3.10+5.2*cos(200)}, {0.40+5.2*sin(200)} )
  -- cycle}

% % Zero set follows the new polygonal edge
% \def\ZeroSetPath{%
%   (2.90,1.05) -- (3.10,0.40) -- (2.20,-1.40) -- (-0.40,-1.80)}

% define \rho position so same in all panels
\pgfmathsetmacro{\rhoX}{0.5}  % x
\pgfmathsetmacro{\rhoY}{0.2}  % y

% -------------------- (a) TOP PANEL --------------------
\begin{subfigure}[t]{0.95\linewidth}
\centering
\begin{tikzpicture}[line cap=round,line join=round,>=Latex, font=\small]
\path[use as bounding box] (-2.6,-2.2) rectangle (3.6,2.6);

\filldraw[blob] \BlobPath;

% \rho_\psi
\coordinate (psi) at (3.10,0.40); 
\node[point] at (psi) {};
\node[anchor=west] at (psi) {$\psi$};

% % iso fidelity semi circles, full circles clipped by 'blobpath'
% \begin{scope}
%   \clip \BlobPath;
%   \foreach \r in {0.5,1.0,1.5,2.0,2.5,3.0, 3.5, 4, 4.5, 5, 5.5, 6, 6.5} {
%     \draw[iso] (psi) circle (\r);
%   }
% \end{scope}

% iso fidelity semi circles, full circles clipped by 'blobpath'
\begin{scope}
  \clip \BlobPath;
  \foreach \r in {0.5,2.2,3.9} {
    \draw[iso] (psi) circle (\r);
  }
\end{scope}

% arrow and label
\coordinate (rho) at (\rhoX,\rhoY);
\node[point] at (rho) {};
\node[anchor=south] at (rho) {$\rho$};
\draw[->,thick] (psi) -- (rho)
  node[midway,sloped,above=2pt,fill=none,inner sep=1pt,text=black,align=center] {$\mathcal{R}_\psi(\rho)=$\\$-\ln\langle\psi\vert\rho\vert\psi\rangle$};

% % free operations arrow and label
% \coordinate (rhoA) at (-0.2,0.8);
\coordinate (rhoB) at (-0.6,-0.8);
% \node[point] at (rhoA) {};
% \node[point] at (rhoB) {};
% \draw[<->,thick,dashed] (rhoA) -- (rhoB)
%   node[midway,left, fill=white, inner sep=1pt] {$\mathcal{R}_\psi$ equal};

\def\req{3.9} 
\begin{scope}
  \clip \BlobPath; % keep it inside the blob
  \draw[<->,thick,dashed,shorten <=2pt,shorten >=2pt]
    ($(psi)+(90:\req)$) arc (90:270:\req)
    node[midway,sloped,below=-4pt,fill=none,rotate=180,inner sep=1pt] {$\mathcal{R}_\psi$ equal};
\end{scope}

% free operations and label (reduced rugosity)
\begin{scope}
  \clip \BlobPath;
  \coordinate (rhoBnext) at (1.25,-0.80);
  % \node[point] at (rhoBnext) {};
  \draw[->, dashed, thick] (rhoB) -- (rhoBnext)
    node[midway, below=2pt, fill=none, inner sep=1pt] {\hspace{2mm}$\mathcal{R}_\psi$ reduced};
\end{scope}

% labels
\node[anchor=west] at (1.8,-1.8) {$\mathcal{R}_{\psi}(\psi)=0$};

\end{tikzpicture}
\caption{Section \ref{sec:generalizing_quantum_state_texture}}
\label{fig:HS-a}
\end{subfigure}

% -------------------- (b) MIDDLE PANEL --------------------
\begin{subfigure}[t]{0.95\linewidth}
\centering
\begin{tikzpicture}[line cap=round,line join=round,>=Latex, font=\small]
\path[use as bounding box] (-2.6,-2.2) rectangle (3.6,2.6);

% Hilbert space
\filldraw[blob] \BlobPath;

% Corners on the boundary
\coordinate (Z1) at (-0.40,-1.80);
\coordinate (Z2) at (2.20,-1.40);
\coordinate (Z3) at (3.10,0.40);
\coordinate (bottomLeftCorner) at ({3.10+5.2*cos(200)},{0.40+5.2*sin(200)});

% Additional points on edge using interpolation
\coordinate (Z4) at ($(bottomLeftCorner)!0.3!(Z1)$); % 30% of the way between z1 and z3

%Example points: bottom three corners (polytope vertices)
\coordinate (phi1) at (Z1);
\coordinate (phi2) at (Z2);
\coordinate (phi3) at (Z3);
\coordinate (phi4) at (Z4);

% free set polytope
\begin{scope}
  \clip \BlobPath;
  \filldraw[fill=black!10, draw=black, line width=1.2pt, dash pattern=on 5pt off 3pt]
    (phi4) -- (phi1) -- (phi2) -- (phi3) -- cycle;
\end{scope}

% Mark and label free set
\node[zeropt] at (phi1) {};
\node[zeropt] at (phi2) {};
\node[zeropt] at (phi3) {};
\node[zeropt] at (phi4) {};
\node[anchor=west] at ($(phi1)+(-0.25,-0.28)$) {$\psi_{1}$};
\node[anchor=west] at ($(phi2)+(-0.08,-0.28)$) {$\psi_{2}$};
\node[anchor=west] at ($(phi3)+(0.08,-0.08)$) {$\psi_{3}$};
\node[anchor=west] at ($(phi4)+(-0.25,-0.28)$) {$\psi_{4}$};

% Target state
\coordinate (rho) at (\rhoX,\rhoY);
\node[point] at (rho) {};
\node[anchor=south] at (rho) {$\rho$};

% nearest point (set arbitrarily between 1 and 3
\coordinate (pmin) at ($(phi4)!0.5!(phi3)$);

% minimum arrow
\node[point] at (pmin) {}; % mark the midpoint
\draw[->,thick,black] (pmin) -- (rho)
  node[midway,sloped,above=2pt,fill=none,inner sep=1pt]{min};

% Rugosity equation
\node[align=left,fill=none,inner sep=2pt] at (0.45,1)
  {$\underline{\mathcal{R}}_{\mathcal{F}}(\rho)=-\ln\displaystyle\max_{\sigma\in\mathcal{F}}F(\rho,\sigma)$};

% free set label
\node[align=left,fill=none,inner sep=2pt] at (1,-1)
  {$\mathcal{F}$};

\end{tikzpicture}
\caption{Section \ref{sec:relationship_with_measures}}
\label{fig:HS-b}
\end{subfigure}

% -------------------- (c) BOTTOM PANEL --------------------
\begin{subfigure}[t]{0.95\linewidth}
\centering
\begin{tikzpicture}[line cap=round,line join=round,>=Latex, font=\small]
\path[use as bounding box] (-2.6,-2.2) rectangle (3.6,2.6);

% Hilbert space
\filldraw[blob] \BlobPath;

% USE SAME Z DEFINITIONS AS PANEL B
% % Corners on the boundary
% \coordinate (Z0) at (2.90,1.05);
% \coordinate (Z1) at (-0.40,-1.80);
% \coordinate (Z3) at (2.20,-1.40);
% \coordinate (Z5) at (3.10,0.40);

%Example points: bottom three corners (polytope vertices)
\coordinate (phi1) at (Z1);
\coordinate (phi2) at (Z2);
\coordinate (phi3) at (Z3);

% free set polytope
\begin{scope}
  \clip \BlobPath;
  \filldraw[fill=black!10, draw=black, line width=1.2pt, dash pattern=on 5pt off 3pt]
    (phi1) -- (phi2) -- (phi3) -- cycle;
\end{scope}

% Mark and label free set
\node[zeropt] at (phi1) {};
\node[zeropt] at (phi2) {};
\node[zeropt] at (phi3) {};
\node[anchor=west] at ($(phi1)+(-0.25,-0.28)$) {$\psi_{1}$};
\node[anchor=west] at ($(phi2)+(-0.08,-0.28)$) {$\psi_{2}$};
\node[anchor=west] at ($(phi3)+(0.08,-0.08)$) {$\psi_{3}$};

% Target state
\coordinate (rho) at (\rhoX,\rhoY);
\node[point] at (rho) {};
\node[anchor=south] at (rho) {$\rho$};

% nearest point (set arbitrarily between 1 and 3
\coordinate (pmin) at ($(phi1)!0.45!(phi3)$);

% minimum arrow
\node[point] at (pmin) {}; % mark the midpoint
\draw[->,thick,black] (pmin) -- (rho)
  node[midway,sloped,above=2pt,fill=none,inner sep=1pt]{min};

% Rugosity equation
\node[align=left,fill=none,inner sep=2pt] at (0.45,1)
  {$\underline{\mathcal{R}}_{\mathcal{F}_{o}}(\rho)=-\ln\displaystyle\max_{\sigma\in\mathcal{F}_{o}}F(\rho,\sigma)$};

% Orthogonality label
\node[align=left,fill=none,inner sep=2pt] at (-0.8,0.2)
  {$\langle\psi_{i}\vert\psi_{j}\rangle= \delta_{ij}$};

% fixed point condition
\node[align=center,fill=none,inner sep=2pt] at (-0.8,-0.7)
  {$\Lambda(\sigma) = \sigma$,\\ $\newline\sigma\in \mathcal{F}_{o}$};

% free set label
\node[align=left,fill=none,inner sep=2pt] at (1.8,-1)
  {$\mathcal{F}_{o}$};

\end{tikzpicture}
\caption{Section \ref{sec:fixed}}
\label{fig:HS-c}
\end{subfigure}

%\vspace{6mm}

\caption{
\justifying
Illustrations of texture generalizations. (a) Single zero-resource state, with concentric curves marking equal fidelity (equal rugosity), covered in Sec.~\ref{sec:generalizing_quantum_state_texture}. (b) Extension to a convex free set, with rugosity lower-bounded by the maximum fidelity to this set, covered in Sec.~\ref{sec:relationship_with_measures}. (c) Imposing fixed-point constraints on free operations (free states are invariant when acted on by free operations), with the free set restricted to orthogonal extreme points (pure states in the free set are orthogonal to one another), discussed in Sec.~\ref{sec:fixed}.}
\label{fig:threepanel}
\end{figure}
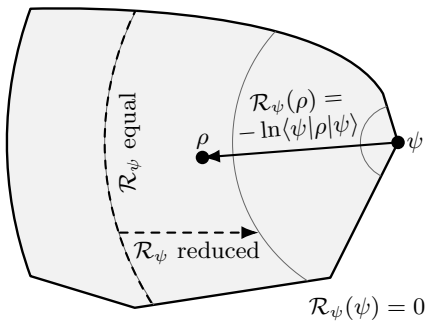
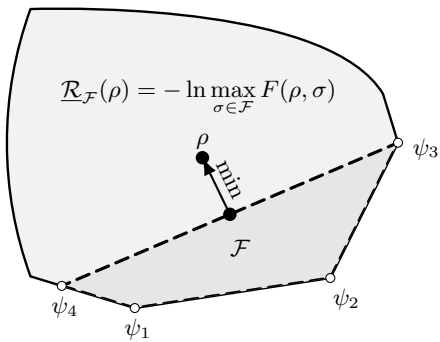
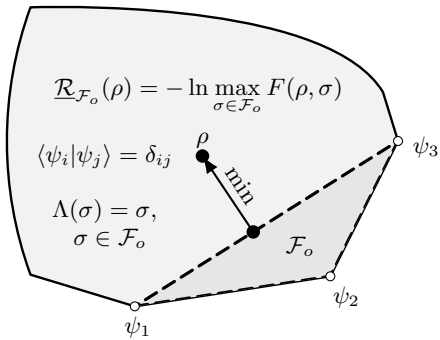

In a recent development \cite{parisio2024quantum}, Parisio introduced the concept of quantum-state texture, a basis-dependent resource theory conveniently quantified by the grand sum of a state's density matrix elements, demonstrating its utility in a novel protocol for identifying controlled-NOT (CNOT) gates. Since the original proposal in \cite{parisio2024quantum}, studies have been released on alternative texture measures \cite{zhang2025quantum,cao2026generalized,muthuganesan2025quantum,wang2025quantifying} and applications in the detection of quantum phase transitions in Ising spin chains \cite{patra2025role}.
Quantum state texture as defined by Parisio is referenced against a single textureless state $\vert f_{1}\rangle=\frac{1}{\sqrt{D}}\sum_{i=1}^{D}\vert i\rangle$ where $\ket{i}$ define the computational basis. The $f_{1}$ state is unique in that, written as a density matrix in the computational basis, all the elements are real and equal. 
It is the equality, or regularity, of the matrix elements of the $f_{1}$ state---the only zero-resource state of the theory---that inspires its moniker ``textureless.''  
A measure of the texture of an arbitrary mixed state $\rho$, or the rugosity, is defined as a function of the fidelity of a state against some zero-resource state $\psi$ normalized for dimension $D$ and is given by (defined in \cite{parisio2024quantum} with $\psi=f_{1}$ but written more generally here in anticipation of our approach)
\begin{equation}
    \mathcal{R}_{\psi}(\rho)=-\ln\frac{\Sigma_{\psi}(\rho)}{D}=-\ln\langle \psi\vert \rho \vert \psi\rangle,
    \label{eq:Rugosity}
\end{equation}
with $\Sigma_{\psi}(\rho)$ defined as
\begin{equation}
    \Sigma_{\psi}(\rho)=D\langle \psi\vert \rho\vert\psi\rangle.
    \label{eq:Sigma_General}
\end{equation}
%and 
%when $\psi=f_{1}$. 
Conveniently, the $\Sigma_{\psi}$ term corresponds to the grand sum of the density matrix elements when $\psi=f_{1}$ \cite{parisio2024quantum},
\begin{equation}
    \Sigma_{f_{1}}(\rho)=D\langle f_{1}\vert\rho\vert f_{1}\rangle=\sum_{i,j}\langle i\vert \rho\vert j\rangle.
    \label{eq:grandsum}
\end{equation}
Parisio then showed that by preparing identical but random inputs to unknown circuit layers, repeated measurements of the texture reveal the presence of CNOT gates as well as reduce the possible bases of the CNOT gates to four. 
While effective, the original formulation of texture and its associated gate-identification protocol naturally raise two questions: (i)~can the approach be extended beyond a single specific reference state---the textureless state---and (ii)~is the use of the ``grand sum," leveraged initially for its simplicity, strictly necessary to perform gate identification and other tasks related to texture? Resolving these questions could provide important insight into the experimental practicality of the gate identification protocol.

In this work, we recast the gate identification protocol in terms of estimating the average fidelity to an \textit{arbitrary pure state $\psi$} without any reference to either the grandsum or the textureless state.
We then demonstrate that, using the same procedure as the original protocol, we can still on average distinguish CNOT and single qubit gates for all but a set of basis states of zero measure on the Bloch sphere, where averaged measurement outcomes on qubits that have underwent a CNOT operation are identical to those that have not. 
In other words, the general $\Sigma_{\psi}$ (and corresponding $\mathcal{R}_{\psi}$) defined above can be used to
achieve gate identification with no particular reference to $f_{1}$ or the grand sum.
Extending beyond this application, we then consider versions of quantum state texture where any pure state can be chosen as the minimum resource state. In doing so, we prove monotonicity and other resource-theoretic requirements without relying on the grand sum shortcut \cite{parisio2024quantum}.
We note that fidelity-based constructions are ubiquitous and natural tools across quantum resource theories. For instance, in the resource theory of magic, the stabilizer fidelity \cite{bravyi2019simulation} and min-relative entropy of magic \cite{liu2022many} are bound classical simulation costs and quantify nonstabilizerness. In our context, however, the core contribution is not the use of fidelity as a novel metric; rather, refocusing the study of state texture back onto fidelity allows us to generalize the resource beyond its original formulation.

A visual summary of the major components of this resource theory is shown in Fig.~\ref{fig:threepanel}\subref{fig:HS-a}, where the single minimum resource (pure) state $\rho_{\psi}$ lies on the boundary of the space with remaining states organized into isofidelity (hence equal-resource-measure) slices; the states of highest resource sit on the furthest edge of the space. 
More generally, this theory is sufficiently rich to subsume or---with extensions to minimum resource sets as depicted in Fig.~\ref{fig:threepanel}\subref{fig:HS-b}---reproduce measures of established resources such as coherence and imaginarity, offering a unified perspective on basis-dependent quantum resources \cite{parisio2024quantum,wang2025quantifying}.
Finally, we introduce a family of so-called ``fixed-point theories'' [shown in Fig.~\ref{fig:threepanel}\subref{fig:HS-c}] that connects seemingly disparate resources such as state texture, genuine coherence \cite{de2016genuine} and athermality \cite{PhysRevLett.111.250404}. 
For these latter two scenarios, where a single free resource state $\psi$ is replaced with a set $\mathcal{F}$ ($\mathcal{F}_{o}$), we quantify the resource through either the direct convex-roof extension of the single-state measure—denoted $\mathcal{R}_{\mathcal{F}}$ ($\mathcal{R}_{\mathcal{F}{o}}$)—or its lower bound, $\underline{\mathcal{R}}_{\mathcal{F}}$ ($\underline{\mathcal{R}}_{\mathcal{F}_{o}}$). We emphasize here that the monotonicity properties of these measures under our defined free operations differ, with $\mathcal{R}_{\psi}$, $\underline{\mathcal{R}}_{\mathcal{F}}$, and $\underline{\mathcal{R}}_{\mathcal{F}_{o}}$ obeying weak monotonicity and the convex roof extensions $\mathcal{R}_{\mathcal{F}}$ and $\mathcal{R}_{\mathcal{F}_{o}}$ violating strong monotonicity as found through explicit examples.

\section{Gate Identification}

Identifying an unknown gate is of general interest in quantum information science, for which process tomography is the most general approach \cite{o2004quantum}. 
However, process tomography is often demanding or unnecessary if only distinguishability between gates is desired \cite{acin2001statistical,duan2009perfect}, which is an important task for characterizing quantum systems and relates closely to the entangling power of a specific gate set \cite{chefles2005entangling}. 
In this section, we rederive the main components of the gate identification protocol of \cite{parisio2024quantum} but from a more general perspective of the expressions $\mathcal{R}_{\psi}$ and $\Sigma_{\psi}$. Although the original treatment selected $\psi=f_{1}$, we will find that the selection of $\psi$ is mostly irrelevant to whether the protocol succeeds. 

\begin{figure*}

\def\w{\ar@{-}[l]}
\def\W{\ar@{=}[l]}
\def\WControl{\ar@[color3bg]@{=}[l]}
\def\WSingle{\ar@[color2bg]@{=}[l]}
%%%%%%%%%%%%%%%%%%%%%%%%%%%%%%%%%%%%%%%%%%%%%%%%%%%%%%%%%%%%%%%%%%%%%%%%%%%%%
% labels

% simple label
\def\A#1{\save []="#1" \restore}

%%%%%%%%%%%%%%%%%%%%%%%%%%%%%%%%%%%%%%%%%%%%%%%%%%%%%%%%%%%%%%%%%%%%%%%%%%%%%
% single qubit operations

\def\op#1{*+[F]{\rule[-0.2ex]{0ex}{2.1ex}#1}}	% operator in box
\def\b{*={\bullet}}
\def\o{*={\oplus}}
\def\t{*={\times}}				% for swap gate
\def\sq{*=<6pt,6pt>[F]{}}			% square, for controlled-phase
\def\m#1{\left[\matrix{#1}\right]}		% matrix shortcut
\def\z{*+[]{\rule[-0.2ex]{0ex}{2.1ex}~|0\>}}	% re-init to |0>
\def\discard{*[]{\rule[-0.2ex]{0.75pt}{2.1ex}~}}	% vertical ``|''
\def\slash{*={/}}				% slash for wire bundles

%%%%%%%%%%%%%%%%%%%%%%%%%%%%%%%%%%%%%%%%%%%%%%%%%%%%%%%%%%%%%%%%%%%%%%%%%%%%%
% nop's

\def\N{*-{}\W}
\def\NControl{*-{}\WControl}
\def\NSingle{*-{}\WSingle}
\def\n{*-{}\w}

%%%%%%%%%%%%%%%%%%%%%%%%%%%%%%%%%%%%%%%%%%%%%%%%%%%%%%%%%%%%%%%%%%%%%%%%%%%%%
% misc definitions

\def\>{\rangle}
\def\<{\langle}
\def\ua{\uparrow}

% measurement box
\def\meter{*+[]{\put(-3,0){\includegraphics[scale=.5]{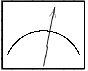}}~~~~}%
		\ar@{-}[l]}

%%%%%%%%%%%%%%%%%%%%%%%%%%%%%%%%%%%%%%%%%%%%%%%%%%%%%%%%%%%%%%%%%%%%%%%%%%%%%
% qubit names (and also revert to qubit wires, vs, cbit wires)

\def\q#1{*+{\rule[-0.2ex]{0ex}{2.1ex}|#1\>}}
\def\qv#1#2{*+{\rule[-0.2ex]{0ex}{2.1ex}|#1\>=|#2\>}}
	
%%%%%%%%%%%%%%%%%%%%%%%%%%%%%%%%%%%%%%%%%%%%%%%%%%%%%%%%%%%%%%%%%%%%%%%%%%%%%
% multiple qubit gates

\newbox{\xyqcircSbox}

\def\gspace#1{*+{\rule[-0.2ex]{0ex}{2.1ex}%
    \setbox\xyqcircSbox=\hbox{$#1$}%
    \hspace*{\wd\xyqcircSbox}}}
	
% n-qubit operation #1=box label, #2=number of qubits (eg d=2 qubits, ddd=4)
\def\gnqubit#1#2{\gspace{#1}
		 \save [].[#2]!C="qq"*[F]\frm{}\restore
		 \save "qq"*[]{#1} \restore}

% two-qubit operation
\def\gtwo#1{\gnqubit{#1}{d}}

% two-qubit operation
\def\gthree#1{\gnqubit{#1}{dd}}

%%%%%%%%%%%%%%%%%%%%%%%%%%%%%%%%%%%%%%%%%%%%%%%%%%%%%%%%%%%%%%%%%%%%%%%%%%%%%
% ``D'' style measurement gate a-la-cleve, at Michael Nielsen's request

\def\dmeterwide#1#2{*{\xy <0pt,-8pt>;<0pt,8pt> **@{-};
		    <0pt,-8pt>;<#2,-8pt> **@{-} ;
		    <0pt, 8pt>;<#2, 8pt> **@{-} ;
		    <#2,0pt>-<5pt,0pt>*{#1} ;
		    <#2,0pt>*\cir<8pt>{r_l}\endxy}}

\def\meter{*+<0pt,12pt>[]{\includegraphics[scale=.5]{./graphics/meter.eps}}%
        \ar@{-}[l]}
%%%%%%%%%%%%%%%%%%%%%%%%%%%%%%%%%%%%%%%%%%%%%%%%%%%%%%%%%%%%%%%%%%%%%%%%%%%%%

% definitions for _old_ the circuit elements
% \def\gAxA{*-{}\w\A{gAxA}}
% \def\gAxB{*-{}\w\A{gAxB}}
% \def\gAxC{*-{}\w\A{gAxC}}
% \def\gAxD{*-{}\w\A{gAxD}}
% \def\gAxE{*-{}\w\A{gAxE}}
% \def\gBxA{\op{\hspace{10pt}}\w\A{gBxA}}
% \def\gBxB{\op{\hspace{10pt}}\w\A{gBxB}}
% \def\gBxC{\op{\hspace{10pt}}\w\A{gBxC}}
% \def\gCxA{\gnqubit{\hspace{10pt}}{d}\w\A{gCxA}}
% \def\gCxB{\gspace{U}\w\A{gCxB}}
% \def\gDxA{*-{}\w\A{gDxA}}
% \def\gDxB{*-{}\w\A{gDxB}}
% \def\gCxC{*-{}\w\A{gCxC}}
% \def\gBxD{*-{}\w\A{gBxD}}
% \def\gBxE{*-{}\w\A{gBxE}}
% \def\gDxC{\gnqubit{\hspace{10pt}}{d}\w\A{gDxC}}
% \def\gDxD{\gspace{U}\w\A{gDxD}}
% \def\gExA{\op{\hspace{10pt}}\w\A{gExA}}
% \def\gExB{\op{\hspace{10pt}}\w\A{gExB}}
% \def\gExC{\op{\hspace{10pt}}\w\A{gExC}}
% \def\gCxE{\op{\hspace{10pt}}\w\A{gCxE}}
% \def\gFxA{\op{\hspace{10pt}}\w\A{gFxA}}
% \def\gGxA{*-{}\w\A{gGxA}}
% \def\gFxB{*-{}\w\A{gFxB}}
% \def\gGxB{*-{}\w\A{gGxB}}
% \def\gFxC{*-{}\w\A{gFxC}}
% \def\gGxC{*-{}\w\A{gGxC}}
% \def\gExD{*-{}\w\A{gExD}}
% \def\gFxD{*-{}\w\A{gFxD}}
% \def\gGxD{*-{}\w\A{gGxD}}
% \def\gDxE{*-{}\w\A{gDxE}}
% \def\gExE{*-{}\w\A{gExE}}
% \def\gFxE{*-{}\w\A{gFxE}}
% \def\gGxE{*-{}\w\A{gGxE}}
% \def\gHxA{\op{U}\w\A{gHxA}}
% \def\gHxB{\op{U}\w\A{gHxB}}
% \def\gHxC{\op{U}\w\A{gHxC}}
% \def\gHxD{\op{U}\w\A{gHxD}}
% \def\gHxE{\op{U}\w\A{gHxE}}
% \def\gIxA{\meter\w\A{gIxA}}
% \def\gIxB{\meter\w\A{gIxB}}
% \def\gIxC{\meter\w\A{gIxC}}
% \def\gIxD{\meter\w\A{gIxD}}
% \def\gIxE{\meter\w\A{gIxE}}

\def\gAxA{*-{}\w\A{gAxA}}
\def\gAxB{*-{}\w\A{gAxB}}
\def\gAxC{*-{}\w\A{gAxC}}
\def\gAxD{*-{}\w\A{gAxD}}
\def\gAxE{*-{}\w\A{gAxE}}
\def\gBxA{\op{\hspace{10pt}}\w\A{gBxA}}
\def\gBxB{\op{\hspace{10pt}}\w\A{gBxB}}
\def\gBxC{\op{\hspace{10pt}}\w\A{gBxC}}
\def\gCxA{\b\w\A{gCxA}}
\def\gCxB{\o\w\A{gCxB}}
\def\gDxA{*-{}\w\A{gDxA}}
\def\gDxB{*-{}\w\A{gDxB}}
\def\gCxC{*-{}\w\A{gCxC}}
\def\gBxD{*-{}\w\A{gBxD}}
\def\gBxE{*-{}\w\A{gBxE}}
\def\gDxC{\gnqubit{\hspace{10pt}}{d}\w\A{gDxC}}
\def\gDxD{\gspace{U}\w\A{gDxD}}
\def\gExA{\op{\hspace{10pt}}\w\A{gExA}}
\def\gExB{\op{\hspace{10pt}}\w\A{gExB}}
\def\gExC{\op{\hspace{10pt}}\w\A{gExC}}
\def\gCxE{\op{\hspace{10pt}}\w\A{gCxE}}
\def\gFxA{\op{\hspace{10pt}}\w\A{gFxA}}
\def\gGxA{*-{}\w\A{gGxA}}
\def\gFxB{*-{}\w\A{gFxB}}
\def\gGxB{*-{}\w\A{gGxB}}
\def\gFxC{*-{}\w\A{gFxC}}
\def\gGxC{*-{}\w\A{gGxC}}
\def\gExD{*-{}\w\A{gExD}}
\def\gFxD{*-{}\w\A{gFxD}}
\def\gGxD{*-{}\w\A{gGxD}}
\def\gDxE{*-{}\w\A{gDxE}}
\def\gExE{*-{}\w\A{gExE}}
\def\gFxE{*-{}\w\A{gFxE}}
\def\gGxE{*-{}\w\A{gGxE}}
\def\gHxA{\op{U}\w\A{gHxA}}
\def\gHxB{\op{U}\w\A{gHxB}}
\def\gHxC{\op{U}\w\A{gHxC}}
\def\gHxD{\op{U}\w\A{gHxD}}
\def\gHxE{\op{U}\w\A{gHxE}}
\def\gIxA{\meter\w\A{gIxA}}
\def\gIxB{\meter\w\A{gIxB}}
\def\gIxC{\meter\w\A{gIxC}}
\def\gIxD{\meter\w\A{gIxD}}
\def\gIxE{\meter\w\A{gIxE}}
% definitions for bit labels and initial states

\def\bA{ \q{\phi}}
\def\bB{ \q{\phi}}
\def\bC{ \q{\phi}}
\def\bD{ \q{\phi}}
\def\bE{ \q{\phi}}

\def\entangleHist{\begin{tikzpicture}
\begin{axis}[
    width=0.35\textwidth,
    height=0.18\textwidth,
    ybar interval,
    ymin=0,
    ymax=0.6,
    xmin=-0.03,
    xmax=2.03,
    xlabel={$\sum_{f_1}(\rho^C) = 2\langle f_1|\rho^C | f_1 \rangle$},
    ylabel={Frequency},
    xtick={0,0.5,1,1.5,2},
    grid=none,
    tick label style={font=\scriptsize},
    label style={font=\small},
    axis line style={draw=color3bg},
]
\addplot+[fill=gray!45, draw=black] coordinates {
    (1.0,0.446)
    (1.2,0.185)
    (1.4,0.142)
    (1.6,0.120)
    (1.8,0.107)
    (2.0,0)
};
\addplot[sharp plot, red, dashed, thick, mark=none]
coordinates {(1.3514,0) (1.3514,0.6)};
\end{axis}
\end{tikzpicture}}

\def\sepHist{\begin{tikzpicture}
\begin{axis}[
    width=0.35\textwidth,
    height=0.18\textwidth,
    ybar interval,
    grid=none,
    ymin=0,
    ymax=0.6,
    xmin=-0.03,
    xmax=2.03,
    xlabel={$\sum_{f_1}(\rho^S) = \sum_{f_1}(\rho^T) =  2\langle f_1|\rho^S | f_1 \rangle$},
    ylabel={Frequency},
    xtick={0,0.5,1,1.5,2},
    tick label style={font=\scriptsize},
    label style={font=\small},
    axis line style={draw=color2bg},
]
\addplot+[fill=gray!45, draw=black] coordinates {
    (0.0,0.1) (0.2,0.1) (0.4,0.1) (0.6,0.1) (0.8,0.1)
    (1,0.1) (1.2,0.1) (1.4,0.1) (1.6,0.1) (1.8,0.1) (2.0,0.0)
};
\addplot[sharp plot, red, dashed, thick, mark=none]
coordinates {(1.0,0) (1.0,0.6)};
\end{axis}
\end{tikzpicture}}
% The quantum circuit and output histograms as two side-by-side xymatrix objects.
% This keeps the circuit rows independent of the histogram rows, so the \vdots
% row is not stretched by the plot height.

% \newcommand{\CircuitXyMatrix}{%
% \xymatrix@R=5pt@C=10pt{
%     \bA & \gAxA &\gBxA &\gCxA &\gDxA &\gExA &\gFxA &\gGxA &\gHxA &\gIxA &\N  
% \\  \bB & \gAxB &\gBxB &\gCxB &\gDxB &\gExB &\gFxB &\gGxB &\gHxB &\gIxB &\N  \\
% % \\  \bC & \gAxC &\gBxC &\gCxC &\gDxC &\gExC &\gFxC &\gGxC &\gHxC &\gIxC &\N  
% % \\  \bD & \gAxD &\gBxD &\n   &\gDxD &\gExD &\gFxD &\gGxD &\gHxD &\gIxD &\N 
% \\ \vdots & & & & \vdots & & & & \vdots & 
% \\  \bE & \gAxE &\gBxE &\gCxE &\gDxE &\gExE &\gFxE &\gGxE &\gHxE &\gIxE &\N 
% %
% % Vertical lines and other post-xymatrix latex
% %
% \save "gBxA"."gFxE"!C *+<32pt,30pt>[F]\frm{} \restore
% }}

\newcommand{\CircuitXyMatrix}{%
\xymatrix@R=5pt@C=8pt{
    \bA & \gAxA &\gBxA & \n &\gCxA &\gDxA  &\gGxA &\gHxA &\gIxA &\NControl  
\\  \bB & \gAxB &\gBxB & \n &\gCxB &\gDxB   &\gGxB &\gHxB &\gIxB &\N  \\
% \\  \bC & \gAxC &\gBxC &\gCxC &\gDxC &\gExC &\gFxC &\gGxC &\gHxC &\gIxC &\N  
% \\  \bD & \gAxD &\gBxD &\n   &\gDxD &\gExD &\gFxD &\gGxD &\gHxD &\gIxD &\N 
\\ \vdots & & & & \vdots & & & & \vdots & 
\\  \bE & \gAxE &\gFxA & \n &\gFxA &\gExE &\gGxE &\gHxE &\gIxE &\NSingle
%
% Vertical lines and other post-xymatrix latex
%
\save "gBxA"."gExE"!C *+<32pt,30pt>[F]\frm{} \restore
\ar@{-}"gCxA";"gCxB"
\save "gDxA"."gGxA"!C+(-1,2.5) *+<0pt,0pt>{\scriptstyle \mbox{$\rho^C$}} \restore
\save "gDxB"."gGxB"!C+(-1,2.5) *+<0pt,0pt>{\scriptstyle \mbox{$\rho^T$}} \restore
\save "gFxA"."gExE"!C+(7,2.5) *+<0pt,0pt>{\scriptstyle \mbox{$\rho^S$}} \restore
}}

\newcommand{\HistogramXyMatrix}{%
\xymatrix@R=3pt@C=0pt{
    \\
    \entangleHist \\
    \\
    \sepHist
}}
\begin{tabular}{@{}c@{\hspace{1.5em}}c@{}}
(a) & (b)\\
$\CircuitXyMatrix$ & $\HistogramXyMatrix$
\end{tabular}%
% -----------------------------------------------------------------------------
% Illustrative output histograms for Parisio's gate-identification protocol.
% Replace the coordinates below with measured/simulated data from your protocol.
% The horizontal axis is written as a generic identification score S, where
% larger S indicates stronger evidence for an entangling gate.
% -----------------------------------------------------------------------------

    \caption{\justifying (a) Schematic illustration of the gate-identification protocol proposed in Ref.~\cite{parisio2024quantum}. 
The input to the unknown $N$-qubit circuit is a product state 
$|\phi\rangle^{\otimes N}$, where $|\phi\rangle$ is sampled uniformly according 
to the Haar measure. The unknown portion of the circuit is indicated by the 
boxed region. After the circuit is applied, each output qubit is measured 
locally in a basis containing the reference textureless state $|f_1\rangle$; 
this measurement is represented by a basis-changing unitary $U$ followed by a computational-basis measurement. The control, target, and ``single'' output qubits, where ``single'' indicates a qubit that has undergone only a single-qubit operation, are represented by $\rho^{C}$, $\rho^{T}$, and $\rho^{S}$, respectively. 
(b) Illustrative histograms of the estimated grand sum $\Sigma_{f_1}$ over \textit{many} Haar-random input states, comparing a \textit{control} qubit influenced by an entangling gate (top) with a qubit affected only by nonentangling operations (bottom). The separation between these distributions provides the statistical signature used to identify which regions of the circuit contain entangling gates. Note that while the statistical distribution for purely single-qubit operations is invariant to the specific gates applied, the plotted distribution for the CNOT control qubit represents the idealized case where the entangling gate acts in the exact basis of the measurement apparatus. In this aligned case, the target qubit $\rho^{T}$ produces a distributions identical to that of the single qubit $\rho^{S}$, as indicated by the axis label on the bottom histogram. The analytical average for the control qubit is $\overline{\Sigma_{f_1}(\rho^{C})} = 4/3$, while the invariant single-qubit operations yield $\overline{\Sigma_{f_1}(\rho^{S})} = 1$. Average values $\overline{\Sigma_{f_1}(\rho^{C})}$ and $\overline{\Sigma_{f_1}(\rho^{S})}$ have been labeled with vertical red dashed lines.}
% \jml{I love this figure! My remaining recommendations: (i) Change $\ket{\psi}$ to $\ket{\phi}$ for consistency with the text's uses of $\psi$ for resourceless states. (ii) Change the $x$-axis $\braket{f_1|\rho_c|f_1}$ to the grand sum $\sum_{f_1}(\rho_c)$ since this is the specific quantity proposed for the protocol. I believe this will in turn change the domain from$[0,1]$ to $[0,2]$. (iii) Show the explicit averages---I think $\overline{\sum_{f_1}(\rho_c)}=\frac{4}{3}$ and $\overline{\sum_{f_1}(\rho_s)}=1$---either in the figure or in the caption. (iv) Make the two plots consistent as to whether they contain dashed gridlines.}\bk{These should now be incorporated}}
    \label{fig:gate_identification}
\end{figure*}

Figure \ref{fig:gate_identification}(a) depicts the gate identification protocol described in this section. Following the original protocol of Ref.~\cite{parisio2024quantum}, we consider a ``single layer'' of quantum operations. While each qubit is subjected to at most one entangling event, these operations are defined with respect to an unknown local single-qubit product basis, hence, what we identify as a CNOT may effectively be a standard CNOT dressed by local single-qubit rotations. Our task is to distinguish between purely single-qubit gates and these basis-shifted CNOT gates without performing full state tomography, relying only on identical but random input qubits. 

By evaluating measurement expectation values over many Haar-random input states, the protocol generates distinct probability density functions, visualized as the histograms in Fig.~\ref{fig:gate_identification}(b). These illustrative histograms are derived assuming the CNOT gate is exactly in the basis of the measurement apparatus (so the unknown local single-qubit rotations are the identity in Fig.~\ref{fig:gate_identification}), resulting in a probability density function for the grand sum of $f(\Sigma) = \frac{1}{2\sqrt{\Sigma-1}}$ for $\Sigma \in (1, 2)$ on the control qubit, and $f(\Sigma)=\frac{1}{2}$ for $\Sigma \in (0,2)$ on a qubit that experiences only nonentangling operations.  Note that this perfectly aligned scenario represents the most extreme case, yielding the sharpest statistical signature. Since the actual gate basis is unknown, the full protocol requires measuring in two complementary bases to guarantee the signature is captured regardless of orientation. Ultimately, comparing the mean of any observed distribution against the uniform single-qubit baseline provides the statistical evidence necessary to detect the presence of entangling operations and importantly, the locations of control and target qubits.
% \jml{Is the single-layer restriction part of Parisio's original protocol or an additional constraint in our example? How the text is currently written, it sounds like the latter, but I seem to recall the former being true.} \lqt{The assumption mentioned in the footnote should be in the figure caption, since the assumption is used to generate the pdf in the figure.}\jml{I think the information in the footnote is sufficiently important that it should be be in the main text.}\bk{All of these comments should now be incorporated}

With this background in mind, we now generalize the CNOT identification protocol by replacing the textureless state $f_1$ with an arbitrary reference $\psi$.
Taking   $\overline{\Sigma_\psi(\rho)}$ as the quantity of interest---i.e., \cref{eq:Sigma_General} averaged over Haar-random inputs---we first observe that
\begin{equation}
    \overline{\Sigma_\psi(\rho^S)}=1
\end{equation}
for the output $\rho^S$ after single-qubit gates only---in agreement with \cite{parisio2024quantum} and following from the fact that the mean fidelity between a fixed but arbitrary pure state and pure state chosen randomly according to the Haar measure is $1/2$ \cite{zyczkowski2005average}.

We next consider this quantity when a CNOT gate has acted in an unknown local single-qubit product basis.
The input state in this case is two identical but Haar-random input qubits.
Note that this is not equivalent to choosing a single $D=4$ state randomly, nor is it the same as choosing both single qubits randomly. It is this distinction that underpins the difference in measurement statistics compared to the single-qubit gate case. 
An arbitrary qubit can be written as $\vert\chi\rangle=\cos\frac{\theta}{2}\vert c\rangle+e^{\imag\phi}\sin\frac{\theta}{2}\vert c'\rangle$ where $c$ and $c'$ form an orthogonal basis, chosen as the (currently unknown) basis of the CNOT gate. 
Note that ``Haar random'' does not mean choosing $\theta$ and $\phi$ uniformly but rather choosing them such that we fairly sample the Bloch sphere. For simplicity, define $a=\cos\frac{\theta}{2}$ and $b=e^{\imag\phi}\sin\frac{\theta}{2}$. 

The input state for two qubits is
\begin{equation}
\vert\chi\rangle\otimes\vert\chi\rangle=a^{2}\vert cc\rangle+ab\vert cc'\rangle+ba\vert c'c\rangle+b^{2} \vert c'c'\rangle,
\end{equation}
which the action of the CNOT gate transforms into
\begin{equation}
\text{CNOT}\vert\chi\rangle\otimes\vert\chi\rangle=a^{2}\vert cc\rangle +ab\vert cc'\rangle +b^{2}\vert c'c\rangle +ba\vert c'c'\rangle.
\end{equation}
We can then trace each qubit out since we will be observing their statistics independently as 
\begin{widetext}
\begin{equation}
    \rho^{C}=\vert a\vert^{2}\vert c\rangle \langle c\vert +(a^{2}b^{*2}+\vert a\vert^{2}\vert b\vert^{2})\vert c\rangle\langle c'\vert +(a^{*2}b^{2}+\vert a\vert^{2}\vert b\vert^{2})\vert c'\rangle\langle c\vert +\vert b\vert^{2}\vert c'\rangle\langle c'\vert
\end{equation}
and
\begin{equation}
    \rho^{T}=(\vert a\vert^{4}+\vert b\vert^{4})\vert c\rangle \langle c\vert +(\vert a\vert^{2}ab^{*}+\vert b\vert^{2}a^{*}b)\vert c\rangle\langle c'\vert +(\vert a\vert^{2}a^{*}b+\vert b\vert^{2}ab^{*})\vert c'\rangle\langle c\vert +2\vert a\vert^{2}\vert b\vert^{2}\vert c'\rangle\langle c'\vert,
\end{equation}
where $C$ denotes the ``control'' and $T$ the ``target'' qubit, respectively. 

We now consider the mean value of $\Sigma_{\psi}$ for both the target and control qubits. Direct substitution gives
\begin{equation}
    \begin{aligned}
        \Sigma_{\psi}(\rho^{C})&=2\left[\vert a\vert^{2}\vert\langle \psi \vert c\rangle\vert^{2} +(a^{2}b^{*2}+\vert a\vert^{2}\vert b\vert^{2})\langle \psi \vert c\rangle\langle c' \vert \psi\rangle +(a^{*2}b^{2}+\vert a\vert^{2}\vert b\vert^{2})\langle \psi \vert c'\rangle\langle c \vert \psi\rangle +\vert b\vert^{2}\vert\langle \psi \vert c'\rangle\vert^{2}\right]\\
        \Sigma_{\psi}(\rho^{T})&=2\left[(\vert a\vert^{4}+\vert b\vert^{4})\vert\langle \psi \vert c\rangle\vert^{2} +(\vert a\vert^{2}ab^{*}+\vert b\vert^{2}a^{*}b)\langle \psi \vert c\rangle\langle c' \vert \psi\rangle +(\vert a\vert^{2}a^{*}b+\vert b\vert^{2}ab^{*})\langle \psi \vert c'\rangle\langle c \vert \psi\rangle +2\vert a\vert^{2}\vert b\vert^{2}\vert\langle \psi \vert c'\rangle\vert^{2}\right].\\
    \end{aligned}
\end{equation}
\end{widetext}
From these we can then calculate the mean $\Sigma_{\psi}$ using the fact that the input states are Haar random and hence $\mathbb{E}[\vert a\vert^{2}]=\mathbb{E}[\vert b\vert^{2}]=1/2$ and $\mathbb{E}[\vert a\vert^{2}\vert b\vert^{2}]=1/6$, while many of the other terms average to zero \cite{parisio2024quantum}:
\begin{equation}
    \begin{aligned}
        \overline{\Sigma_{\psi}(\rho^{C})}
        &=1+\frac{2}{3}\text{Re} \langle \psi \vert c\rangle\langle c' \vert \psi\rangle\\
        \overline{\Sigma_{\psi}(\rho^{T})}&=\frac{2}{3}\left[1+\vert\langle \psi \vert c\rangle\vert^{2}\right].\\
    \end{aligned}
    \label{eq:Sigma}
\end{equation}
Deviation of either $\overline{\Sigma_{\psi}(\rho^{C})}$ or $\overline{\Sigma_{\psi}(\rho^{T})}$ from one indicates the presence of a CNOT gate. 
More information is revealed by measuring in the basis $\vert\psi_{H}\rangle=H\vert\psi\rangle$ where $H$ is the Hadamard gate (in the lab basis, not in the CNOT basis).
For this we use $\Sigma_{\psi_{H}}(\rho)=2\langle\psi_{H}\vert \rho\vert \psi_{H}\rangle$, and the average expressions for the control and target qubits then become
\begin{equation}
    \begin{aligned}
        \overline{\Sigma_{\psi_{H}}(\rho^{C})}&=1+\frac{2}{3}\text{Re} \langle \psi_{H} \vert c\rangle\langle c' \vert \psi_{H}\rangle \\
        \overline{\Sigma_{\psi_{H}}(\rho^{T})}&=\frac{2}{3}\left[1+\vert\langle \psi_{H} \vert c\rangle\vert^{2}\right]\\
    \end{aligned}
    \label{eq:Sigma_Hadamard}
\end{equation}
We note that Eqs.~\eqref{eq:Sigma} and \eqref{eq:Sigma_Hadamard} together fully constrain the CNOT basis~\cite{parisio2024quantum}, since $\psi$ and $\psi_{H}$ are in the lab basis and explicitly prepared by us.

By inspection, if the chosen free state $\psi$ happens to align with the CNOT basis such that all expressions in Eqs.~\eqref{eq:Sigma} and \eqref{eq:Sigma_Hadamard} equal unity, the results would be indistinguishable from those of single-qubit gates, and the gate identification procedure would  fail to detect the true presence of the CNOT. 
To determine the possibility and measure of such ``pathological'' free states $\psi$, we set $ \overline{\Sigma_{\psi}(\rho^{C})} =  \overline{\Sigma_{\psi}(\rho^{T})} =  \overline{\Sigma_{\psi_{H}}(\rho^{C})} = \overline{\Sigma_{\psi_{H}}(\rho^{T})} = 1$ in the above and solve for $\psi$:
\begin{equation}
    \begin{aligned}
        \text{Re} \langle \psi \vert c\rangle\langle c' \vert \psi\rangle =\text{Re} \langle \psi_{H} \vert c\rangle\langle c' \vert \psi_{H}\rangle =0\\
        \vert\langle \psi \vert c\rangle\vert^{2}=\vert\langle \psi_{H} \vert c\rangle\vert^{2}=\frac{1}{2}.
    \end{aligned}
\end{equation}

By parameterizing $\psi$ in the unknown CNOT basis as
$\vert\psi\rangle=\cos\frac{\theta}{2}\vert c\rangle+e^{\imag\phi}\sin\frac{\theta}{2}\vert c'\rangle$ where $\theta\in[0,\pi]$ and $\phi\in[0,2\pi)$, we find two possible solutions
\begin{equation}
\vert\psi_{\pm}\rangle=\frac{1}{\sqrt{2}}\left(\vert c\rangle\pm \imag\vert c'\rangle\right)
\end{equation}
We can now enforce the final constraint that these two solutions be connected in the lab basis by a Hadamard transformation. In other words, we want to find rotations $U$ such that 
\begin{equation}
HU\vert\psi_{-}\rangle=U\vert\psi_{+}\rangle.
\label{eq:hupsi_relation}
\end{equation}
We explicitly solve for this $U$ in Appendix \ref{App:Rotation} and find
\begin{equation}
    U=e^{\imag\mu}R_{y}\left(\frac{\pi}{4}\right)R_{z}(\nu_{2}),
\end{equation}
where $R_y$ and $R_z$ denote rotations about the $y$ and $z$ axes of the Bloch sphere. This $U$ defines a set of states for which the gate identification fails: meaning the single-qubit unitary rotations are indistinguishable from CNOT gates through statistical measures of $\Sigma_\psi$ and $\Sigma_{\psi_H}$.  This family of states has one nontrivial degree of freedom, $\nu_{2}$, and traces out a great circle. Hence, the family of failure states has zero measure on the surface of the Bloch sphere, and therefore the probability of selecting such a state at random is exactly zero in the continuous, uniform limit.

\begin{figure}[t!]
\begin{center}
\begin{tikzpicture}[line cap=round, line join=round, >=Triangle]

  %------------------ Parameters ------------------%
  \def\R{2.2}       % sphere radius
  \def\eqr{0.85}    % projected equator minor radius
  \def\Tilt{50}     % visual tilt for Ry(pi/4)
  \def\Nu{35}       % position of U psi_+ along failure circle (deg)

  % Colors
  \colorlet{cnot}{black}
  \colorlet{lab}{blue!90!white}
  \colorlet{fail}{green!70!black}
  \colorlet{lamc}{violet!80!black}

  %------------------ Canvas ------------------%
  \clip(-3.4,-3.2) rectangle (3.6,3.8);

  % Sphere
  \draw (0,0) circle (\R cm);

  % CNOT equator (reference, dashed gray)
  \draw[dash pattern=on 3pt off 3pt, draw=black!40] (0,0) ellipse (\R cm and 1.1*\eqr cm);

  %------------------ CNOT basis axes & labels (black) ------------------%
  \draw[cnot,->,thick] (0,0) -- (0,0,\R);
  \draw[cnot,->,thick] (0,0) -- (0,\R,0);      % +z_c = |c>
  \draw[cnot,->,thick] (0,0) -- (\R,0,0);      % +x_c
  \node[anchor=south] at (0,\R+0.12) {\scriptsize $\ket{c}$};
  % \node[anchor=north] at (0,-\R-0.12) {\scriptsize $\ket{c'}$};

  % λ± in the CNOT basis: ±y on the equator
  \fill[lamc] (0,0, \R) circle (1.6pt);
  \fill[lamc] (0,0,-\R) circle (1.6pt);
  \node[lamc,anchor=west] at (0,0,-1.1*\R)  {\scriptsize $\psi_{+}$};
  \node[lamc,anchor=west] at (0,0,0.8*\R)  {\scriptsize $\psi_{-}$};

  %------------------ Lab axis (blue): U|c>, U|c'> ------------------%
  % Lab z-axis is CNOT z tilted by Ry(pi/4) toward +x
  \path let \p1 = ({\R*cos(\Tilt)},{\R*sin(\Tilt)}) in
    coordinate (labZ) at (\x1,\y1)
    coordinate (labZminus) at (-\x1,-\y1);
  \draw[lab,->,thick] (0,0) -- (labZ);
  \node[lab,anchor=west] at ($(labZ)+(0.12,0)$) {\scriptsize $U\ket{c}$};
  % \node[lab,anchor=east] at ($(labZminus)+(-0.12,0)$) {\scriptsize $U\ket{c'}$};

  % Wedge indicating the fixed tilt Ry(pi/4)
  \draw[fill=gray!40, fill opacity=0.15, draw=gray!70]
    (0,0) -- (90:\R*0.75) arc[start angle=90,end angle=\Tilt,radius=\R*0.75] -- cycle;
  \node at (65:1.9) {\scriptsize $R_y(\tfrac{\pi}{4})$};

  %------------------ Failure great circle (green) ------------------%
  % This is the image of the CNOT equator under U; it is the equator
  % of the LAB frame, i.e., a great circle perpendicular to the lab z-axis.
  % To reflect that visually, rotate the ellipse by (Tilt + 90°) so its plane
  % is orthogonal to the drawn lab axis direction.
  \def\eqrLab{1.4*\eqr} % this is to make up for the fact that we're using the perspective of the axis being 'down and to the left' instead of 'out of the page'
  \begin{scope}[rotate around={\Tilt+90:(0,0)}]
    \draw[fail,very thick,dash pattern=on 6pt off 3pt]
      (0,0) ellipse (\R cm and \eqrLab cm);

    % Place U psi_± as antipodal points along this circle at angle \Nu
    \pgfmathsetmacro{\ux}{\R*cos(\Nu)}
    \pgfmathsetmacro{\uy}{\eqrLab*sin(\Nu)}
    \fill[fail]  (\ux,\uy) circle (1.9pt);
    \fill[fail]  (-\ux,-\uy) circle (1.9pt);

    % Labels for U states
    \node[fail,anchor=west] at ($( \ux,\uy)-(-0.2,0.28)$)  {\scriptsize $U\psi_{+}$};
    \node[fail,anchor=east] at ($(-\ux,-\uy)+(0.14,0.00)$) {\scriptsize $U\psi_{-}$};

    % ν2 arrow showing motion along the failure circle
    \def\ArcA{\Nu-30}
    \def\ArcB{\Nu}
    \draw[->,fail,thick]
      ({1.1*\R*cos(\ArcA)},{\eqrLab*sin(\ArcA)}) arc[start angle=\ArcA, end angle=\ArcB,
        x radius=1.5*\R, y radius=1.5*\eqrLab];
    \node[fail] at ({\R*cos(\Nu-15)+0.35},{\eqrLab*sin(\Nu-15)+0.18}) {\scriptsize $\nu_{2}$};
  \end{scope}

  % Origin
  \draw[fill] (0,0) circle (1.4pt);

\end{tikzpicture}
\caption{
\justifying
Black axes define the CNOT basis with one basis ket $\ket{c}$ listed ($\ket{c'}$ omitted for clarity) and $\psi_\pm$ on the CNOT equator. $R_{z}(\nu_{2})$ rotates the pair $\psi_{\pm}$ along the CNOT equator (they remain antipodal) followed by a fixed tilt from $R_y(\pi/4)$ (resulting in the green dashed circle). The blue arrow indicates the tilt from the CNOT to the lab basis $U\psi_\pm$, also tilted by the fixed $R_y(\pi/4)$.}
\label{Fig:Gate}
\end{center}
\end{figure}
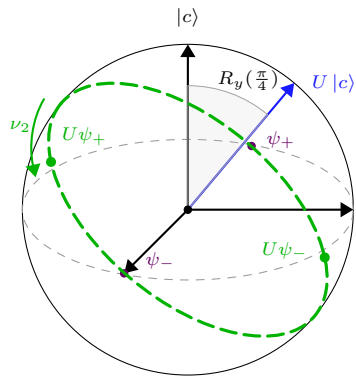

The connection between the two bases, the action of $U$, and the great circle of points for which the two sets of gates are indistinguishable are visualized in Fig. \ref{Fig:Gate}.
In summary, we see that CNOT gates and single-qubit gates can be successfully distinguished for texture-like theories [as defined in Eqs.~(\ref{eq:Rugosity}--\ref{eq:grandsum})] for \emph{any reference pure state $\psi$ except for those lying on a great-circle of measure zero on the Bloch sphere.}
This finding significantly generalizes the original protocol requiring projective measurement onto the strictly prescribed state $f_1$~\cite{parisio2024quantum}, motivating the further generalization of texture-like resource theories in the following sections. 
% \jml{This is not something for the revision, but as I've continued thinking about texture: it seems to me that not only can texture be generalized, but actually the textureless state $\ket{f_1}$ has \emph{no preferred position at all} for gate identification. Whenever the CNOT basis is unknown, $\ket{f_1}$ is literally no better than $\ket{0}$---both have the same chances of achieving a specific overlap with the pathological states on the great circle. Is this interpretation correct?}\bk{Yes, indeed there seems to be nothing about texture that is specific to gate identification, really the only important part about the textureless state as chosen is it allows for the use of the grand sum}

\section{Generalizing Quantum State Texture}
\label{sec:generalizing_quantum_state_texture}
In the previous section, we found that one's choice of the free state $\psi$ has little bearing on whether gate identification is possible, so long as it cannot be written as $\ket{\psi}=\frac{1}{\sqrt{2}}e^{\imag\mu}R_{y}\left(\frac{\pi}{4}\right)R_{z}(\nu_{2})(\ket{c}+\imag\ket{c'})$.
We now consider whether one can equivalently construct a resource theory from any choice of a so-called ``textureless state'' $\psi$. We posit two conditions for this new form of ``generalized texture'': (i) there is a single minimum resource state $\psi$ that serves as a fixed point for free operations $\Lambda^\psi$, and (ii) the measure of resource takes the form of Eq.~\eqref{eq:Rugosity}. Later, we will show how this can be generalized further to recover resource measures of known resources such as genuine coherence \cite{de2016genuine}.
We will begin by showing that we can recover all of the desirable properties of a resource theory for an arbitrary $\psi$ and without the use of the grand sum. 

Following an analogous approach to the resources of entanglement \cite{chen2014comparison} and coherence \cite{baumgratz2014quantifying}, among others, we define a measure $M(\rho)$ as satisfying
\begin{enumerate}
    \item \textit{Nonnegativity}. $M(\rho) \geq 0 \quad \forall \rho$. $M(\rho)=0$ if and only if $\rho$ is free.
    \item \textit{(Weak) Monotonicity}\footnote{In comparison to weak monotonicity, where measurement outcomes are unknown, strong monotonicity averages over each Kraus operator $K_{j}$ such that  $\sum_{j}p_{j}M(\sigma_{j})\le M(\rho)$ where $\sigma_{j}=K_{j}\rho K_{j}^{\dagger}/p_{j}$ and $p_{j}=\text{Tr}(K_{j}\rho K_{j}^{\dagger})$.}. $M(\rho)$ does not increase under free operations; i.e. $M(\Lambda_{\text{free}}(\rho)) \leq M(\rho)$.
\end{enumerate}
In what follows we will establish these features for $\mathcal{R}_{\psi}$.

The free operations for our generalization of state texture admit a Kraus decomposition of the form
\begin{equation}
    \Lambda^{\psi}(\rho)=\sum_{n=1}^NK_{n}^{\psi}\rho (K_{n}^{\psi})^{\dagger}
    \label{eq:basic_channel_operation}
\end{equation}
where $\sum_{n}(K_{n}^{\psi})^{\dagger}K_{n}^{\psi}=\mathds{1}$ and $N$ is the Kraus rank. 
In accordance with condition (i) we assume that states of minimum resource must be fixed, and hence $\Lambda^{\psi}(\rho_{\psi})=\rho_{\psi}$ for the notation $\rho_{\psi}=\vert \psi\rangle\langle \psi\vert$, which is only possible when
\begin{equation}
\label{eq:Kpsi}
    K_{n}^\psi \vert \psi\rangle = a_{n}^{\psi}\vert \psi\rangle
\end{equation}
for some constant $a_{n}^{\psi}$,
because $\ket{\psi}$ is pure and therefore cannot be generated through a convex sum of other states. 
In order for free operations $\Lambda^\psi$ to recover the state $\ket{\psi}$, it must factor out of \cref{eq:basic_channel_operation}.

In Appendix \ref{app:monotonicity_of_generalized_texture}, we prove that this fixed-point condition guarantees $\mathcal{R}_{\psi}$ satisfies both nonnegativity and weak monotonicity [i.e., $\mathcal{R}_{\psi}(\Lambda^{\psi}(\rho)) \leq \mathcal{R}_{\psi}(\rho)$] for arbitrary pure and mixed states, establishing it as a valid resource measure.

We now consider the form of free operations in this resource theory.
The Hilbert space can be decomposed into $\mathcal{H}=\mathcal{H}_{\psi}\oplus\mathcal{H}_{\psi}^{\perp}$, where $\mathcal{H}_{\psi}=\text{span}\{\vert\psi\rangle\}$ and $\mathcal{H}_{\psi}^{\perp}$ is the orthogonal complement with $\dim\{\cH_{\psi}^{\perp}\}=D-1$.
Accordingly we define the basis of the full space as $\{\vert\psi\rangle, \vert\psi_{\perp}^{(2)}\rangle, \vert\psi_{\perp}^{(3)}\rangle,...,\vert\psi_{\perp}^{(D)}\rangle\}$
We want to find free operations that fulfill $\Lambda^{\psi}(\rho_\psi)=\rho_\psi$, which is accomplished for Kraus operators of the form
\begin{equation}
    K^{\psi}_n \;=\;
\begin{pmatrix}
  a_n^{\psi} & T_{n}^{\psi} \\
  0   & L^{\psi}_n,
\end{pmatrix}
\end{equation}
expressed in the $1\oplus(D-1)$ dimensional block basis relative to $\mathcal{H}=\mathcal{H}_{\psi}\oplus\mathcal{H}_{\psi}^{\perp}$. 
Here $a_{n}^{\psi}\in\mathbb{C}$ and we have expanded the $A_{n}$ operator into two constituent parts: $L_n\in\mathbb{C}^{(D-1)\times(D-1)}$ is a Kraus operator acting only on the orthogonal support of $\psi$ and $T_{n}\in\mathbb{C}^{1\times(D-1)}$ is a (one-way) transition operator. 
Note that the trace-preserving condition $\sum_{n}(K_{n}^{\psi})^{\dagger}K_{n}^{\psi}=\mathds{1}$ forces $\sum_{n}\vert a_{n}^{\psi}\vert^{2}=1$, which in turn implies the fixed point condition we desire since 
\begin{equation}
\sum_{n}K_{n}^\psi\vert\psi\rangle\langle\psi\vert (K_{n}^\psi)^{\dagger}=\sum_{n}\vert a_{n}^\psi\vert^{2}\vert\psi\rangle\langle\psi\vert=\vert\psi\rangle\langle\psi\vert.
\end{equation}
The $T_{n}$ operator only couples states one direction from the subspace of resourceful states to the subspace of the free state.
Consider an arbitrary input pure state 
\begin{equation}
    \vert\Psi\rangle=
    \begin{pmatrix}
    c_{f}\\
    \mathbf{c}_{r}
    \end{pmatrix}
\end{equation}
decomposed into $\ket{\Psi}=c_{f}\ket{\psi}+\sum_{i=2}^D c_{r}^{(i)}\ket{\psi_\perp^{(i)}}$.
Then calculating the left hand side of the channel map:
\begin{equation}
K_{n}^\psi\vert\Psi\rangle=    \begin{pmatrix}
a_{n}^{\psi}c_{f}+T_{n}^{\psi}\mathbf{c}_{r}\\
    L_{n}^{\psi}\mathbf{c}_{r}
    \end{pmatrix}.
\end{equation}
The action is analogous to amplitude damping from the resourceful space into the free space. 
Hence, the free operations $\Lambda^\psi$ fix the free resource state $\psi$ while allowing for resource-destroying maps (those that couple from resourceful states to the free state). 

So far, we have considered a single minimum resource state. Other notable resource theories with a single free state include purity, which can equivalently be quantified by a texture measure~\cite{patra2025role}, and athermality \cite{brandao2013resource}, where the Gibbs free-energy state possesses no resource. 
We will revisit these connections in Sec.~\ref{sec:fixed}.

% \lqt{To shorten section III, my suggestion would be to move the content roughly from Eq.(18) to Eq. (23) to the appendix.}\bk{I had to cut a little more because I realized then we used a term only defined in the appendix in the main text.}

\section{Relationship with Measures of other Resources}
\label{sec:relationship_with_measures}
Up to this point, we have considered generalizations of texture from a zero point defined explicitly as $f_{1}$---the textureless state---to any pure state $\psi$. Here we go one step further and, with the addition of a minimization, extend from a single zero-resource \emph{state} to a convex zero-resource \emph{set}. 
Figure~\ref{fig:threepanel}\subref{fig:HS-b} provides a schematic illustration where a zero-resource set comprises a convex set $\cF$ of three pure states $\{\psi_1,\psi_2,\psi_3\}$ (here $\mathcal{F}$ need only be convex with no other constraints;  Sec.~\ref{sec:fixed} will add constraints for free operations). 
In the cartoon we show the lower bound $\underline{\mathcal{R}}_{\mathcal{F}}$ (which we will later show is itself a valid resource measure) rather than the most natural extension of the rugosity through a convex-roof extension (which we will call $\mathcal{R}_{\mathcal{F}}$) given that it has a simple geometric form as the maximum fidelity between the target state and the free set $\mathcal{F}$ . 
In general we will see that $\mathcal{R}_\cF$ has no analytical solution; however, for free sets like the real states or the incoherent states we recover closed form expressions for single qubits (using a formulation similar to that of \cite{patra2025role}).

For the convex set $\mathcal{F}$ of free resource states, we introduce the generalized rugosity of an arbitrary pure state $\phi$ as
\begin{equation}
\label{eq:rugosityPure_in_mixed_comparison}
    \mathcal{R}_{\mathcal{F}}(\phi)=-\ln\max_{\sigma\in\mathcal{F}}\langle\phi\vert\sigma\vert\phi\rangle,
\end{equation}
where we use symbols $\phi$ and $\psi$ for pure states as opposed to $\rho$ and $\sigma$ for mixed states, which should be clear from context going forward.
Using the linearity of $\langle\phi\vert\sigma\vert\phi\rangle$ for a fixed $\phi$ and an expansion of any state $\sigma\in\mathcal{F}$ as $\sigma=\sum_{k}c_{k}\vert\psi_{k}\rangle\langle\psi_{k}\vert$ where $\psi_{k}\in\mathcal{F}$, we find
\begin{equation}
\langle\phi\vert\sigma\vert\phi\rangle=\sum_{k}c_{k}\vert\langle\psi_{k}\vert\phi\rangle\vert^{2}\le\max_{k}\vert\langle\psi_{k}\vert\phi\rangle\vert^{2},
\end{equation}
which is always attainable, and hence, Eq.~\eqref{eq:rugosityPure_in_mixed_comparison} can be written as
\begin{equation}
\label{eq:rugosityPure}
    \mathcal{R}_{\mathcal{F}}(\phi)=-\ln\max_{\psi_{k}\in\mathcal{F}}\vert\langle\phi\vert\psi_{k}\rangle\vert^{2}.
\end{equation}

Now we move to mixed states $\rho=\sum_i c_{i}\vert\phi_{i}\rangle\langle\phi_{i}\vert$ through a convex-roof extension such that the generalized rugosity is the minimum across all possible decompositions. Thus we define
\begin{equation}
    \mathcal{R}_{\mathcal{F}}(\rho)=\min_{c_{i},\phi_{i}}\sum_i c_{i} \mathcal{R}_{\mathcal{F}}(\phi_{i}),
    \label{eq:convex_roof_exact}
\end{equation}
which goes to zero whenever $\rho\in \mathcal{F}$, since it then has a decomposition $\{c_i,\ket{\psi_{i}}\}$ where every $\psi_i\in\mathcal{F}$, driving each term to zero. 

The pure state definition in Eq.~\eqref{eq:rugosityPure_in_mixed_comparison} enjoys the useful interpretation as a fidelity between an input state and the free set $\cF$. 
Unfortunately, the convex-roof extension of Eq.~\eqref{eq:convex_roof_exact} does not have as simple of an interpretation, though it can be bounded by a fidelity-based expression.
We define this lower bound of $\mathcal{R}_{\mathcal{F}}(\rho)$ to be 
\begin{equation}
    \underline{\mathcal{R}}_{\mathcal{F}}(\rho)=-\ln\max_{\sigma\in\mathcal{F}}F(\rho,\sigma),
    \label{eq:r_tilde}
\end{equation}
where $F(\rho,\sigma)=(\text{tr}\sqrt{\sqrt{\rho}\sigma\sqrt{\rho}})^{2}$.
Note that if the state $\rho\in\mathcal{F}$ then $F(\rho,\sigma)=1$ and $ \underline{\mathcal{R}}_{\mathcal{F}}(\rho)=0$.
And when $\rho$ is pure ($\rho=\vert\phi\rangle\langle\phi\vert$), the two definitions agree
\begin{equation}
\begin{aligned}
    \underline{\mathcal{R}}_{\mathcal{F}}(\phi)&=-\ln\max_{\sigma\in\mathcal{F}}F(\vert\phi\rangle\langle\phi\vert,\sigma)\\
    &=-\ln\max_{\sigma\in\mathcal{F}}\langle\phi\vert\sigma\vert\phi\rangle\\
    &=\mathcal{R}_{\mathcal{F}}(\phi).
\end{aligned}
\end{equation}

As defined, $F$ is not jointly concave. However, the logarithm in $\underline{\mathcal{R}}_\cF$ allow us to recast the problem in terms of the root fidelity $\sqrt{F}$, which is jointly concave \cite{watrous2018theory}.
In particular we can write $\underline{\mathcal{R}}_{\mathcal{F}}(\rho)=-2\ln \max_{\sigma\in\mathcal{F}}\sqrt{F(\rho,\sigma)}$.
Given this joint concavity we can use the well known result that partial maximization of a jointly concave function over a convex set preserves concavity \cite{boyd2004convex}, and therefore $\max_{\sigma\in\mathcal{F}}\sqrt{F(\rho,\sigma)}$ is concave.
Finally, since $-\ln(\cdot)$ is nonincreasing and convex the composition here is again convex, allowing us to apply Jensen's theorem.
Specifically,
\begin{equation}
    \underline{\mathcal{R}}_{\mathcal{F}}\left(\sum_{i}c_{i}\vert\phi_{i}\rangle\langle\phi_{i}\vert\right) \le\sum_{i}c_{i}\underline{\mathcal{R}}_{\mathcal{F}}(\phi_{i})
    =\sum_{i}c_{i}\mathcal{R}_{\mathcal{F}}(\phi_{i}),
\end{equation}
which holds for any decomposition, including the minimum:
\begin{equation}
    \underline{\mathcal{R}}_{\mathcal{F}}(\rho)\le\min_{c_{i},\phi_{i}}\sum_{i}c_{i}\mathcal{R}_{\mathcal{F}}(\phi_{i})=\mathcal{R}_{\mathcal{F}}(\rho).
\end{equation}
In Sec.~\ref{sec:fixed} we will show that $\underline{\mathcal{R}}_{\mathcal{F}}$ is itself a resource measure.
Accordingly, the generalized rugosity defined by Eq.~\eqref{eq:convex_roof_exact} is distinct from, yet lower-bounded by, the optimization over the fidelity [Eq.~\eqref{eq:r_tilde}]. In the following subsections we apply our generalized rugosity to two resource theories in particular: imaginarity (Sec.~\ref{sec:imaginarity}) and basis-dependent coherence (Sec.~\ref{sec:basisDepCoh}).
For ease of reference we include Fig. \ref{fig:bloch_text_coh_imag} which illustrates the minimum and maximum resource states for state texture [Fig. \ref{fig:bloch_text_coh_imag}\subref{fig:bloch_min_texture}], coherence [Fig. \ref{fig:bloch_text_coh_imag}\subref{fig:bloch_min_coherence}], and imaginarity [Fig. \ref{fig:bloch_text_coh_imag}\subref{fig:bloch_min_imaginarity}] in the single-qubit case ($D=2$).

Before proceeding, we clarify the scope and role of the single-qubit examples presented in the following subsections. While our generalized resource framework and the convex-roof extension are defined for systems of arbitrary dimension $D$, evaluating the convex roof analytically for mixed states with $D>2$ is computationally demanding and typically requires numerical optimization (as we detail in Appendix \ref{app:numerical}). Therefore, the exact, analytically solvable $D=2$ cases explored in Sections \ref{sec:imaginarity} and \ref{sec:basisDepCoh} are provided specifically as illustrative examples to build physical intuition and concretely demonstrate how our framework recovers known resource measures.

\subsection{Recovering Measures of Imaginarity}
\label{sec:imaginarity}
The resource theory of imaginarity is motivated by fundamental questions about the necessity and operational significance of complex numbers in the formulation of quantum mechanics 
\cite{aleksandrova2013real, wootters2014rebit,wu2021operational,hita2025quantummechanicsbasedreal, hoffreumon2025quantumtheorydoesneed}.
A resource theory of imaginarity can be built on a set of free quantum states consisting of real density matrices in a fixed basis \cite{wu2021resource}.
More specifically, $\mathcal{F}=\{\sigma\in\mathcal{D}(\mathcal{H}):\sigma_{ij} \in \mathbb{R} \;\forall\; i,j\}$ where $\mathcal{D(H)}$ is the set of valid density matrices.
We can find a closed form expression for $\mathcal{R}_{\mathcal{F}}(\rho)$ by expanding any general density matrix as $\rho=\text{Re}\,\rho +\imag\,\text{Im}\,\rho$. Since $\rho$ is Hermitian, $\text{Im}\,\rho$ is skew symmetric ($\text{Im}\,\rho_{ij}=-\text{Im}\,\rho_{ji}$) and $\langle\varphi\vert \text{Im}\, \rho\vert\varphi\rangle=0$ when $\varphi$ is a real vector.
$\mathcal{R}_{\mathcal{F}}(\phi)$ in Eq.~\eqref{eq:rugosityPure} can therefore be simplified to%$
\begin{equation}
\begin{aligned}
\label{eq:rugosityPureReal}
    \mathcal{R}_{\mathcal{F}}(\phi)&=-\ln \max_{\varphi\in\mathcal{F}}\langle\varphi\vert (\text{Re}\,\vert\phi\rangle\langle\phi\vert)\vert\varphi\rangle\\
    &=-\ln \lambda_\mathrm{max}(\text{Re}\, \vert\phi\rangle\langle\phi\vert)\\
\end{aligned}
\end{equation}
where $\lambda_\mathrm{max}(\rho)$ indicates the largest eigenvalue of $\rho$, and this maximization is the well known Rayleigh quotient. Hence, for pure states the rugosity is simply the logarithm of the largest eigenvalue of the real part of the state.

Specializing to single-qubit states [Fig \ref{fig:bloch_text_coh_imag}\subref{fig:bloch_min_imaginarity}] to facilitate visual understanding, we first note 
that the real part of the density matrix for any pure state can be written in the Pauli basis as 
\begin{equation}
    \text{Re }\rho=\frac{1}{2}\left(\mathds{1}+n_{x}\sigma_{x}+n_{z}\sigma_{z}\right)
\end{equation}
for unit-length Bloch vector $\hat{\mathbf{n}}=(n_x,n_y,n_z)$. 
Here we define $r_\mathrm{real}^{2}=n_{x}^{2}+n_{z}^{2}$ such that $r_\mathrm{real}^{2} + n_y^2=1$ for a pure state. Hence, the eigenvalues of interest are given by $\lambda(\text{Re }\rho)=\frac{1}{2}(1\pm\vert r_\mathrm{real}\vert)$~\cite{bruning2012parametrizations} 
and the rugosity follows as
\begin{equation}
\begin{aligned}
    \mathcal{R}_{\mathcal{F}}(\phi)&=-\ln\left(\frac{1+\vert r_\mathrm{real}\vert}{2}\right).
\end{aligned}
\end{equation}

We can now use this result to extend to mixed states via the minimization over decompositions. For a single qubit, any mixed state can be written as the convex sum of two pure states, since geometrically any point inside the Bloch sphere lies on a chord connecting two points on the surface. These two states need not be orthogonal, unlike in the spectral decomposition.
Moreover, we will find that the convex-roof inequality can be saturated with such a two-state, generally nonorthogonal, decomposition, and therefore we restrict to two states in what follows.
Thus, $\rho$ for a qubit can always be written as $\rho=p\vert\phi_{q}\rangle\langle\phi_{q}\vert+(1-p)\vert\phi_{s}\rangle\langle\phi_{s}\vert$ which can be expressed using the Bloch vectors $\hat{q}$ and $\hat{s}$ of each pure state as 
\begin{equation}
\label{eq:decomp}
\rho=\frac{1}{2}\left(\sigma_{0}+\left[p\hat{q}+(1-p)\hat{s}\right]\cdot\hat{\sigma}\right).
\end{equation}
The minimization problem then required to find $\mathcal{R}_{\mathcal{F}}(\rho)$ is given by
\begin{widetext}
\begin{equation}
\label{eq:minimum}
    \mathcal{R}_{\mathcal{F}}(\rho)=\min_{p,\hat{q},\hat{s}}\left[-p\ln\left(\frac{1+\sqrt{1-q_{y}^2}}{2}\right)-(1-p)\ln\left(\frac{1+\sqrt{1-s_{y}^2}}{2}\right)\right],
\end{equation}
where we have leveraged $q_\mathrm{real}^2+q_y^2=s_\mathrm{real}^2+s_y^2=1$ by normalization. 

We first observe that $-\ln(\cdot)$ has a positive second derivative for $0\le x\le 1$ and hence the function is convex, allowing for the application of Jensen's inequality:
\begin{equation}
\label{eq:Jensen}
    \mathcal{R}_{\mathcal{F}}(\rho)\ge-\ln\left[\frac{p}{2}\left(1+\sqrt{1-q_{y}^2}\right)+\frac{1-p}{2}\left(1+\sqrt{1-s_{y}^2}\right)\right].
\end{equation}
\end{widetext}
Saturation occurs when the arguments  of the logarithms are equal, so we look for a decomposition $q_y^2=s_y^2$ subject to  
$pq_\text{real}+(1-p)s_\text{real}=r_\text{real}$ as implied by Eq.~\eqref{eq:decomp}.

To see that this decomposition always exists we begin by analyzing the 
$q_{y}=s_{y}$ solution to $q_{y}^{2}=s_{y}^{2}$ (the  $q_{y}=-s_{y}$ results in a valid decomposition but does not minimize the rugosity). 
Setting $q_{y}=s_{y}=r_{y}$ means that we are guaranteed to recover the $y$ component of the state regardless of mixing value $p$. 
Geometrically, we now have two states $\hat{q}$ and $\hat{s}$ lying along a circle on a plane parallel to the real $xz$ plane of the Bloch sphere, but shifted from by the amount $y=r_{y}$. 
These two states must be somewhere on the circle, not within the disc, as they are pure. %states and must lie on the surface of the sphere.
Note that $\sqrt{1-r_{y}^{2}}\ge|r_\text{real}|$, meaning the radius of the fixed circle in the $xz$ plane is greater than or equal to the remaining components of the state due to $\vert \mathbf{r}\vert^{2}\le1$ (i.e., the state we are reproducing could be mixed). 
Since the disc is the convex hull of this circle, we can reach any point within it with a convex combination of two antipodal pure states on the surface such that the chord connecting them passes through the point of interest and where movement along this chord is mediated by the mixing term $p$.
The special case of the decomposition reducing to a single state occurs for $p\in\{0,1\}$ when $\vert \mathbf{r}\vert^{2}=1$.

Accordingly, the decomposition with $q_y^2=s_y^2$ gives the analytical solution to the imaginarity as 
\begin{equation}
    \mathcal{R}_{\mathcal{F}}(\rho)=-\ln\left(\frac{1+\sqrt{1-r_{y}^{2}}}{2}\right).
\end{equation}
In other words, the imaginarity resource depends only on the $y$ component of the single qubit and mirrors known results \cite{wu2021resource,du2025quantifying}.

\subsection{Recovering Measures of Coherence}
\label{sec:basisDepCoh}

Off-diagonal terms of a density matrix in a fixed basis imply the superposition of states in that basis and enable quantum interference effects.
The resource theory of quantum coherence aims to characterize and quantify this effect \cite{baumgratz2014quantifying, streltsov2017colloquium}.
In this section we are concerned with a basis-dependent notion of coherence, similar to the imaginarity---to be contrasted with the \emph{basis-independent} theory of genuine coherence discussed later in \cref{sec:fixed}, which refers specifically to the more restrictive definition of free operations expressed in Eq.~\eqref{eq:free_ops} below~\cite{de2016genuine}.  
For the present fixed-basis coherence resource theory, the free set is given by $\mathcal{F}=\{\sigma\in\mathcal{D}(\mathcal{H}):\sigma = \sum_i c_i\ket{i}\bra{i}\}$ for a fixed computational basis $i\in\{1,...,D\}\equiv[D]$. 
The pure state rugosity is then found by expanding the state $\phi$ in the computational basis: $\vert\phi\rangle=\sum_i \ket{i}\braket{i|\phi}$. We can therefore find $\mathcal{R}_{\mathcal{F}}(\phi)$ in Eq.~\eqref{eq:rugosityPure} through inspection as the pure states in $\mathcal{F}$ are the basis states in the computational basis:
\begin{equation}
    \mathcal{R}_{\mathcal{F}}(\phi)=-\ln\max_{i\in[D]}|\braket{i|\phi}|^{2}.
\end{equation}
This result is a special case of that found in \cite{zhu2017coherence} and is consistent with other approaches \cite{zhang2020numerical}. 

\begin{figure*}[t!]
        \begin{center}
        \begin{subfigure}[t]{0.3\textwidth}
            \begin{tikzpicture}[line cap=round, line join=round, >=Triangle]
              \clip(-2.19,-2.49) rectangle (2.66,2.58);
              \draw [shift={(0,0)}, lightgray, fill, fill opacity=0.1] (0,0) -- (56.7:0.4) arc (56.7:90.:0.4) -- cycle;
              \draw [shift={(0,0)}, lightgray, fill, fill opacity=0.1] (0,0) -- (-135.7:0.4) arc (-135.7:-33.2:0.4) -- cycle;
              \draw(0,0) circle (2cm);
              \draw [rotate around={0.:(0.,0.)},dash pattern=on 3pt off 3pt] (0,0) ellipse (2cm and 0.9cm);
              \draw (0,0)-- (0.70,1.07);
              \draw [->] (0,0) -- (0,2);
              \draw [->] (0,0) -- (-0.81,-0.79);
              \draw [->] (0,0) -- (2,0);
              \draw [dotted] (0.7,1)-- (0.7,-0.46);
              \draw [dotted] (0,0)-- (0.7,-0.46);
              \draw (-0.08,-0.3) node[anchor=north west] {$\varphi$};
              \draw (0.01,0.9) node[anchor=north west] {$\theta$};
              \draw (-1.5,-0.72) node[anchor=north west] {$\mathbf {\ket{D}}$};
              \draw (2.07,0.3) node[anchor=north west] {$\mathbf{\ket{R}}$};
              \draw (-0.3,2.6) node[anchor=north west] {$\mathbf {|0\rangle}$};
              \draw (0.4,1.65) node[anchor=north west] {$|\psi\rangle$};
              \scriptsize
              \draw [fill] (0,0) circle (1.5pt);

              \draw [fill,red] (-0.81,-0.79) circle (1.5pt);
              \draw [fill,blue] (0.81,0.79) circle (1.5pt);
              \draw [fill] (0.7,1.1) circle (0.5pt);
            \end{tikzpicture}
            \caption{}
            \label{fig:bloch_min_texture}
        \end{subfigure}%
        ~ 
        \begin{subfigure}[t]{0.3\textwidth}
            \begin{tikzpicture}[line cap=round, line join=round, >=Triangle]
              \clip(-2.19,-2.49) rectangle (2.66,2.58);
              \draw(0,0) circle (2cm);
              \draw [rotate around={0.:(0.,0.)},dash pattern=on 3pt off 3pt] (0,0) ellipse (2cm and 0.9cm);
              \draw [rotate around={90.:(0.,0.)},red,thick,pattern=north west lines, pattern color=red] (0,0) ellipse (2cm and 0.9cm);
              \draw [->] (0,0) -- (0,2);
              \draw [->] (0,0) -- (-0.81,-0.79);
              \draw [->] (0,0) -- (2,0);
              % \draw [blue,thick] (-2,0) -- (2,0);
              \draw [fill,blue] (-2,0) circle (1.5pt);
              \draw [fill,blue] (2,0) circle (1.5pt);
              \draw (-1.5,-0.72) node[anchor=north west] {$\mathbf {\ket{D}}$};
              \draw (2.07,0.3) node[anchor=north west] {$\mathbf{\ket{R}}$};
              \draw (-0.3,2.6) node[anchor=north west] {$\mathbf {|0\rangle}$};
              \draw (-0.3,-2) node[anchor=north west] {$\mathbf {|1\rangle}$};
              \scriptsize
              \draw [fill] (0,0) circle (1.5pt);
            \end{tikzpicture}
            \caption{}
            \label{fig:bloch_min_imaginarity}
        \end{subfigure}
        ~
        \begin{subfigure}[t]{0.3\textwidth}
            \begin{tikzpicture}[line cap=round, line join=round, >=Triangle]
              \clip(-2.19,-2.49) rectangle (2.66,2.58);
              \draw(0,0) circle (2cm);
              \draw [rotate around={0.:(0.,0.)},blue,thick] (0,0) ellipse (2cm and 0.9cm);
              \draw [->] (0,0) -- (0,2);
              \draw [->] (0,0) -- (-0.81,-0.79);
              \draw [->] (0,0) -- (2,0);
              \draw [red,thick] (0,-2) -- (0,2);
              \draw (-1.5,-0.72) node[anchor=north west] {$\mathbf {\ket{D}}$};
              \draw (2.07,0.3) node[anchor=north west] {$\mathbf{\ket{R}}$};
              \draw (-0.3,2.6) node[anchor=north west] {$\mathbf {|0\rangle}$};
              \draw (-0.3,-2) node[anchor=north west] {$\mathbf {|1\rangle}$};
              \scriptsize
              \draw [fill] (0,0) circle (1.5pt);
            \end{tikzpicture}
            \caption{}
            \label{fig:bloch_min_coherence}
        \end{subfigure}%
    \end{center}
    \begin{tikzpicture}
            \matrix [draw, above right] at (-8,20) {
         \draw[pattern=north west lines, pattern color=blue,draw=blue] (0,0.1) rectangle (0.25,0.35); &  \node[blue] {Maximum Resource States}; \\
         \draw[pattern=north west lines, pattern color=red,draw=red] (0,0.1) rectangle (0.25,0.35); &  \node[red] {Minimum Resource States}; \\
        };
    \end{tikzpicture}
    \caption{Illustration of minimum (red) and maximum (blue) resource states for single-qubit implementations of (a) quantum state texture, (b) imaginarity, and (c) coherence. Definitions: $\ket{D}\equiv\frac{1}{\sqrt{2}}(\ket{0}+\ket{1})$ and $\ket{R}\equiv\frac{1}{\sqrt{2}}(\ket{0}+\imag\ket{1})$.}
    \label{fig:bloch_text_coh_imag}
\end{figure*}
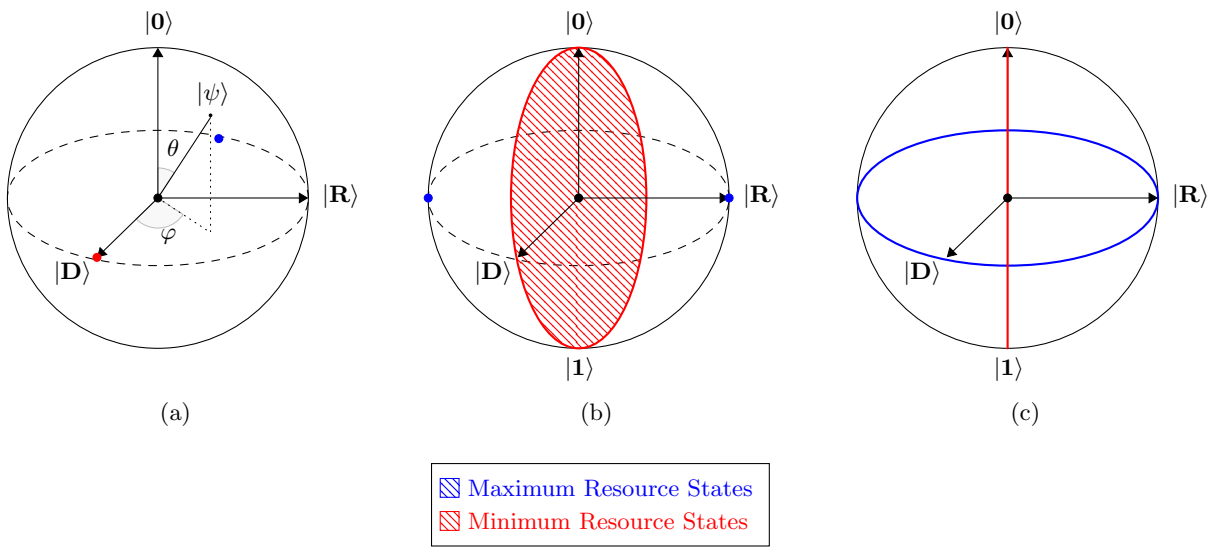

In general for mixed states the convex roof over this is still required; however, for a single qubit [Fig \ref{fig:bloch_text_coh_imag}\subref{fig:bloch_min_coherence}] an analytical solution is possible. The pure state solution can be written in terms of the $z$ Bloch vector component: 
\begin{equation}
    \mathcal{R}_{\mathcal{F}}(\phi)=-\ln\left(\frac{1+\vert n_{z}\vert}{2}\right).
\end{equation}

By applying a minimization procedure and saturation of Jensen's inequality parallel to that used for imaginarity (detailed explicitly in Appendix \ref{app:coherence_derivation}), the convex-roof extension is given by 
\begin{equation}
    \mathcal{R}_{\mathcal{F}}(\rho)=-\ln\left(\frac{1+\sqrt{1-r_{\perp}^{2}}}{2}\right).
\end{equation}
Therefore, as expected, the transverse (perpendicular to $z$) components of the state determine the coherence. In other words, $\mathcal{R}_{\mathcal{F}}(\rho)$ is a function of how far $\rho$ is from the incoherent states (which form the $z$ axis). 

% \lqt{Section IV is already pretty clear and concise, though there is a noticeable parallel between subsections A and B. Perhaps one can go from Eq (46) directly to Eq(48) in subsection B, given that the same derivation procedure is used.}\bk{good idea, I've done exactly this}

%%%%%%%%%%%%%%%
\section{Fixed-Point Resource Theories}
\label{sec:fixed}

In what follows, we modify our generalized theory of texture with two additional constraints on free operations $\Lambda$ that preserve the structure of the resource theory derived in Sec.~\ref{sec:generalizing_quantum_state_texture} and provide a unifying framework for theories such as genuine coherence \cite{de2016genuine}, purity (studied in both the infinitely many~\cite{PhysRevA.67.062104} and few-copy cases~\cite{gour2015resource}), and athermality \cite{brandao2013resource}. 
Our first constraint, given by
\begin{equation}
    \Lambda(\sigma) = \sigma \quad \forall\,\sigma\in \mathcal{F}_{o}, 
    \label{eq:free_ops}
\end{equation}
fixes every zero-resource state $\sigma$ for \textit{all} free operations $\Lambda$ and free set $\mathcal{F}_{o}$. 
In contrast to previous sections where the only restriction in the free set $\mathcal{F}$ was convexity, our second constraint requires all pure states in the free set $\mathcal{F}_{o}$ to be orthogonal (hence the $o$ subscript). 
This orthogonality constraint will force the free operations $\Lambda$ to take a convenient block form, simplifying our analysis, while still containing several known resource theories.
Similarly to the case of a single minimum resource state $\psi$, we may use the same Kraus description of our free operations; i.e. $\Lambda(\rho)=\sum_{n}K_{n}\rho K_{n}^\dagger$. Due to the orthogonality of the pure states in $\mathcal{F}_{o}$, the multipoint case is straightforwardly expressed (in the diagonal $\mathcal{F}_{o}$ basis) 
\begin{equation}
    K_n  \;=\;
\begin{pmatrix}
  \text{diag}(\alpha_{1,n},\alpha_{2,n},...,\alpha_{m,n}) & T_{n} \\
  \mathbf{0}   & L_n
\end{pmatrix}
\label{eq:mp_diagonal_form_free_ops}
\end{equation}
where the Hilbert space $\mathcal{H}=\mathcal{H}_{\mathcal{F}_o}\oplus\mathcal{H}_{\mathcal{F}_{o}}^{\perp}$ is $D$-dimensional and $m$ orthogonal free states comprise $\cF_o$; i.e., $\text{dim}\{\mathcal{H}\}=D$, $\text{dim}\{\mathcal{H}_{\mathcal{F}_{o}}\}=m$, and $\text{dim}(\mathcal{H}_{\mathcal{F}_{o}^{\perp}})=D-m$. 
Trace preservation of the Kraus operators implies that 
\begin{equation}
\begin{aligned}
    \sum_{n}&\vert\alpha_{i,n}\vert^{2}=1\hspace{3mm}\forall\,i\in[m]\\
    \sum_{n}& \text{diag}(\alpha_{1,n},\alpha_{2,n},...,\alpha_{m,n})^{\dagger}T_{n}=0\\
    \sum_{n}&T_{n}^{\dagger}T_{n}+L_{n}^{\dagger}L_{n}=\mathds{1}_{D-m}.\\  
\end{aligned}
\end{equation}
Intuitively, the diagonal nature of the upper left block (corresponding to the free set) means every pure state in the free set is an eigenvector of these operators, while the vanishing lower left block means that nothing can couple out of the free set, and the nonzero upper right corner allows for maps that reduce or destroy resource.
Importantly, if more than one nonorthogonal pure state is a fixed point in this theory we would not be able to diagonalize $K_n$ in this fashion. 

\textit{Definition}.---We shall call any resource theory a \textit{fixed-point resource theory} if any pure states in the free set $\mathcal{F}_{o}$ are orthogonal and the free operations obey Eq.~\eqref{eq:free_ops} and [and hence take the form of Eq.~\eqref{eq:mp_diagonal_form_free_ops}]. Intuitively, we use the term fixed-point resource theory due to the fact that all free states are fixed points to the free operations $\Lambda$. 

We may choose our measure for any fixed-point resource theory to take the form of the logarithmic measure $\mathcal{R}_{\mathcal{F}_o}(\rho)$ and its convex-roof extension defined in Eq.~\eqref{eq:convex_roof_exact}. Indeed, the logarithmic measure is universal as a so-called ``geometric measure" of entanglement; a survey of its properties can be found in \cite{chen2014comparison}. To the best of our knowledge, there is no proof that the convex roof of the logarithmic measure is weakly monotonic, leaving this task as an open question. \cref{app:strong_monotonicity} shows concrete evidence that Eq.~\eqref{eq:convex_roof_exact} violates strong monotonicity. In \cref{app:numerical}, we present numerical results showing that Eq.~\eqref{eq:convex_roof_exact} is likely weakly monotonic.

Nevertheless, one can prove analytically that the \emph{lower bound} $\underline{\mathcal{R}}_{\mathcal{F}_o}(\rho)$ [Eq.~\eqref{eq:r_tilde}] obeys weak monotonicity. In particular, for $\sigma\in\cF_o$ and arbitrary $\rho$,
\begin{equation}
    F(\Lambda(\rho),\Lambda(\sigma))=F(\Lambda(\rho),\sigma)\ge F(\rho,\sigma),
\end{equation}
due to monotonicity of the fidelity $F$ \cite{nielsen2010quantum} and the fixed point condition of the free set $\Lambda(\sigma)=\sigma$.
Since this holds for every $\sigma\in\mathcal{F}_{o}$ we know it also holds for the maximum and hence
\begin{equation}
\begin{aligned}
    \max_{\sigma\in\mathcal{F}_{o}}F(\Lambda(\rho),\sigma)&\ge\max_{\sigma\in\mathcal{F}_{o}}F(\rho,\sigma)\\
    -\ln\max_{\sigma\in\mathcal{F}_{o}}F(\Lambda(\rho),\sigma)&\le-\ln\max_{\sigma\in\mathcal{F}_{o}}F(\rho,\sigma)\\
    \underline{\mathcal{R}}_{\mathcal{F}_o}(\Lambda(\rho))&\le\underline{\mathcal{R}}_{\mathcal{F}_o}(\rho).
\end{aligned}
\end{equation}
Taken together with the fact that $\underline{\mathcal{R}}_{\mathcal{F}_o}(\sigma)=0$ for any state $\sigma\in\mathcal{F}_{o}$ we establish that $\underline{\mathcal{R}}_{\mathcal{F}_o}$ is itself a valid resource measure, as defined in Sec.~\ref{sec:generalizing_quantum_state_texture} by the nonnegativity and monotonicity, rather than merely a lower bound to $\mathcal{R}_{\mathcal{F}_o}$.

These results are analogous to those found for geometric measures of entanglement \cite{chen2014comparison} where convex-roof extensions of logarithmic geometric quantities (similar to our $\mathcal{R}_{\mathcal{F}_o}$) are known to not possess strong monotonicity in general settings, with open questions surrounding their weak monotonicity properties.  In comparison, measures based on the Uhlmann fidelity, similar to our $\underline{\mathcal{R}}_{\mathcal{F}_o}$, are known to exhibit weak monotonicity \cite{chen2014comparison}.

The condition defined in Eq.~\eqref{eq:free_ops} is physically constraining and requires that free operations on the entire free set act effectively as an identity operation. Unsurprisingly, this condition can only be applied to a limited number of existing resource theories. 
Although we recovered a measure of imaginarity from rugosity in Sec.~\ref{sec:imaginarity}, imaginarity \textit{cannot} be framed as a fixed-point resource theory given our definition. This is simply because the pure free states in the resource theory of imaginarity form a set that is not pairwise orthogonal and hence do not fulfill the free-set requirements as defined in our fixed-point resource theory.

An important example of a fixed-point resource theory is that of genuine coherence \cite{de2016genuine}, prominent as a form of coherence where free (genuinely incoherent) operations are laboratory independent, and are thereby connected to entanglement. In this case, the minimum resource states $\sigma\in\mathcal{F}_o$ are defined as those with purely diagonal components:
\begin{equation}
\label{eq:rhoInc}
    \sigma_\mathrm{inc} = \sum_i p_i\ket{i}\bra{i},
\end{equation}
where $\sum_ip_i = 1$. 
Note that this fulfills the orthogonality requirement of the pure states since the only pure states in $\sigma_\text{inc}$ occur for $p_{i}=\delta_{ik}$ and hence are the basis states.
In particular, $\sigma_\text{inc}$ corresponds to the special case where the Kraus operators defined in \cref{eq:mp_diagonal_form_free_ops} are completely diagonal as there is no $L_{n}$ or $T_n$.

The authors of \cite{de2016genuine} define genuinely incoherent free operations $\Lambda_{\text{gi}}(\rho)$ that can only map incoherent states $\sigma_\text{inc}$ to themselves:
\begin{equation}
      \Lambda_{\text{gi}}(\sigma_\mathrm{inc}) = \sigma_\mathrm{inc}.
      \label{eq:genuine_inc_fixed_point}
\end{equation}
which clearly satisfies the fixed point condition of our definition. 
Finally, we note that $\underline{\mathcal{R}}_{\mathcal{F}_o}$ is a weakly monotonic and nonnegative measure of genuine coherence, as we demonstrated for the case of general fixed-point resource theories earlier in this section. 

Despite the complete diagonal form that the Kraus operators take in this case where the free set spans the entire space ($m=D$) there are still nontrivial resource-nonincreasing free operations. In particular, dephasing operations in the diagonal $\mathcal{F}_{o}$ basis leave the diagonal states fixed and hence are free operations. Further, for the special case of complete dephasing this map becomes resource-destroying, for which $\Lambda(\sigma)=\sigma$ for all $\sigma\in\mathcal{F}_{o}$ and $\Lambda(\rho)\in\mathcal{F}_o$ for any resourceful state $\rho\in\mathcal{F}_o^\perp$~\cite{liu2017resource}.
Notably, complete dephasing is not resource-destroying in fixed-point resource theories in general ($m<D$) as not all diagonal states are in the free set in these cases.

Two other special cases of fixed-point resource theories are that of purity \cite{PhysRevA.67.062104,gour2015resource} and athermality \cite{brandao2013resource}. In each case there is a single, mixed, free state in the free set (the latter corresponding to a pure ground state only in the special case of bath temperature $T\to 0$). Hence the orthogonality of the pure states in $\mathcal{F}_{o}$ in each case is trivially satisfied (as there are no pure states in the free set). 
Here the block structure of the free maps changes as the free set is full rank and population transfer between $\mathcal{H}_{\mathcal{F}_{o}^{\perp}}$ and $\mathcal{H}_{\mathcal{F}_{o}}$ through $T_{n}$ is unavailable. 
However, resource-destroying maps of the depolarizing type, which map to a single point, are possible. For example, if the single free state in the free set is given by $\tau$ ($\tau$ is identity for the resource theory of purity and the Gibbs state for athermality) resource destroying maps of the kind $\Lambda(\rho)=\tau$ are possible.

\section{Conclusion and discussion}
In this work, we revisit the role of quantum state texture in a recently proposed protocol to distinguish between CNOT and single-qubit gates in single-layer quantum circuits.
The original protocol \cite{parisio2024quantum} relies on quantification of the quantum resource of state texture, or rugosity---specifically that of Haar-random, but identically prepared, input states following the traversal of the single-layer quantum circuit. 
The protocol requires measuring in more than two bases, determined by the resourceless free state and its respective Hadamard transform, to identify both the location of the CNOT gates and the basis on which the gates act.

We have shown that by generalizing the texture in terms of its fidelity to an arbitrary zero-resource pure state, gate identification can be performed for any pure free state except those lying on a set of measure zero.
Hence, the use of state texture and the accompanying grand sum, as introduced in \cite{parisio2024quantum}, are convenient special cases of a more general gate-identification protocol. %will potentially change
Furthermore, our generalized resource theory of texture is well defined for arbitrary-dimensional Hilbert spaces with a single pure state identified as the zero-resource state, reducing to the original texture result when selecting the zero-texture state $\psi=f_1$ \cite{parisio2024quantum}. 
Although our generalized framework loses the analytical simplicity of the original grand sum, it yields two key benefits for gate identification. Conceptually, it reveals that the protocol's success stems primarily from the diversity of Haar-random input states rather than the mathematical structure of the $f_1$ state. Practically, this approach provides significant experimental flexibility; by proving the protocol functions using nearly any arbitrary reference state, we remove the strict laboratory requirement to prepare and measure in the $f_1$ basis.

Further, one can extend this resource for multiple resourceless states forming a convex set. Critically, this framework allows us to recover single-qubit measures of known resources, such as imaginarity and coherence. 
Further, we define a specific family of resource theories, which we refer to as ``fixed-point'' for which all minimum resource states are unaltered by free operations while maintaining nonnegativity and monotonicity.
In doing so we show that many existing resource theories, including genuine coherence, purity, and athermality, admit realizations within our framework.

The framework we have presented unifies several disparate concepts in the study of quantum resource theories and reevaluates the role of quantum state texture in the gate identification protocol. In future work, one may consider formally identifying the operational significance of fixed-point operations in different contexts to better establish connections between seemingly distinct resources such as athermality, purity, and genuine coherence.

\bibliography{BIB}

\appendix

\renewcommand\thesubfigure{(\alph{subfigure})}

\section{Deriving specific rotations}
\label{App:Rotation}

Before we begin we note the following helpful relationships:
\begin{equation}    \sigma_{y}\vert\psi_{\pm}\rangle=\pm\vert\psi_{\pm}\rangle,\,\,\,\,\sigma_{z}\vert\psi_{\pm}\rangle=\vert \psi_{\mp}\rangle,
\end{equation}
where
\begin{equation}
    \vert\psi_{\pm}\rangle=\frac{1}{\sqrt{2}}\left(\vert c\rangle\pm \imag\vert c'\rangle\right),
\end{equation}
and we emphasize that the Pauli operators act in the CNOT basis. 
Given $\sigma_{z}\vert\psi_{\pm}\rangle=\vert \psi_{\mp}\rangle$ our aim is generally to find $U$ such that $U\sigma_{z}U^{\dagger}=H$. In other words, seek a lab basis where the relationship between the $\psi_{\pm}$ states becomes a Hadamard transformation. 

The most general single-qubit unitary can be decomposed into $U = e^{\imag\mu} R_{z}(\nu_{1}) R_{y}(\tau) R_{z}(\nu_{2})$ where the $R_{k}(\theta)=\sigma_{0}\cos\frac{\theta}{2}-\imag\sigma_k\sin\frac{\theta}{2}$ denotes a rotation around the $k$ axis. In order to solve $U\sigma_{z}U^{\dagger}=H$ we first expand the left-hand side, leveraging the commutativity of $\sigma_z$ and $R_z(\nu_2)$:
\begin{equation}
\label{eq:UZU}
U\sigma_z U^\dagger = \sigma_{z}\cos\tau+\sigma_{x}\cos\nu_{1}\sin\tau+\sigma_{y}\sin\nu_{1}\sin\tau.
\end{equation}
\begin{comment}
\begin{equation}
    \begin{aligned}
        e^{i\mu} R_{z}(\nu_{1}) R_{y}(\tau) R_{z}(\nu_{2}) \sigma_{z}\left(e^{i\mu} R_{z}(\nu_{1}) R_{y}(\tau) R_{z}(\nu_{2})\right)^{\dagger} \\=R_{z}(\nu_{1}) R_{y}(\tau) \sigma_{z} R^{\dagger}_{y}(\tau) R^{\dagger}_{z}(\nu_{1})\\
    \end{aligned}
\end{equation}
where in the first line we used the fact that $e^{i\mu}$ is a phase and that $R_{z}(\nu_{2})$ is a rotation around the $z$ axis and hence commutes with $\sigma_{z}$. 

We can now solve the remaining terms to find
\begin{equation}
    \begin{aligned}
        &R_{z}(\nu_{1}) R_{y}(\tau) \sigma_{z} R^{\dagger}_{y}(\tau) R^{\dagger}_{z}(\nu_{1})\\
        &=\sigma_{z}\cos(\tau)+\sigma_{x}\cos(\nu_{1})\sin(\tau)+\sigma_{y}\sin(\nu_{1})\sin(\tau)\\
    \end{aligned}
\end{equation}
\end{comment}
We next set this equal to $H=\frac{1}{\sqrt{2}}\left(\sigma_{x}+\sigma_{z}\right)$. 
Two solutions for $(\tau,\nu_{1})$ exist, given by $(\frac{\pi}{4},0)$ and $(-\frac{\pi}{4},\pi)$.
However, each of these ultimately trace out an identical set of states when the complete range of input values for $\nu_{2}$ are considered. To see this set the operators formed by each solution equal (up to a global phase)
\begin{equation}
    \begin{aligned}
        U&= U'\\
        R_{y}\left(\frac{\pi}{4}\right) R_{z}(\nu_{2})&= R_{z}(\pi) R_{y}\left(-\frac{\pi}{4}\right) R_{z}(\nu'_{2})\\
        R_{y}\left(\frac{\pi}{4}\right) R_{z}(\nu_{2}-\nu'_{2})&= R_{z}(\pi) R_{y}\left(-\frac{\pi}{4}\right)\\
        R_{y}\left(\frac{\pi}{4}\right) R_{z}(\nu_{2}-\nu'_{2})&= (-\imag)\sigma_{z} R_{y}\left(-\frac{\pi}{4}\right)\\
        R_{y}\left(\frac{\pi}{4}\right) R_{z}(\nu_{2}-\nu'_{2})&= e^{-\frac{\imag \pi}{2}} R_{y}\left(\frac{\pi}{4}\right)\sigma_{z}\\
    \end{aligned}
\end{equation}
where we have indicated the duplicate solution with primes and on the last line used the fact that $\sigma_{z}\sigma_{y}=-\sigma_{y}\sigma_{z}$.
Hence every point reachable by one unitary is reachable by the other (to a global phase, in this case $e^{-\frac{\imag \pi}{2}}$), just with an offset free variable $\nu_{2}=\pi+\nu'_{2}$.
Therefore, the unique set of rotations is given by
\begin{equation}
    U=e^{\imag\mu}R_{y}\left(\frac{\pi}{4}\right)R_{z}(\nu_{2}).
\end{equation}
We have found by construction a family of states where the generalized gate identification protocol fails. This family has one nontrivial degree of freedom, $\nu_{2}$, and traces out a great circle, rendering them of zero-measure on the surface of the Bloch sphere.

\section{Monotonicity of generalized texture}
\label{app:monotonicity_of_generalized_texture}

Starting with \cref{eq:Kpsi}, we define the kernel operators $A_{n}^{\psi}\equiv K_{n}^{\psi}-a_{n}^{\psi}\vert\psi\rangle\langle\psi\vert$. 
Although $A_{n}^{\psi}\vert\psi\rangle=0$, 
$\psi$ is not necessarily in the kernel of $(A_{n}^{\psi})^\dagger$ since Kraus operators are not in general Hermitian~\cite{parisio2024quantum}.
We can now rewrite the fixed zero-point condition as:
\begin{equation}
\begin{aligned}
&\Lambda^{\psi}(\rho_{\psi}) = \sum_n [A_n^\psi + a_n^\psi \rho_{\psi}] \rho_{\psi} [(A_n^\psi)^\dagger + (a_n^\psi)^{*} \rho_{\psi}] = \rho_{\psi} \\
&\implies \sum_n |a_n^\psi|^2 \rho_{\psi} = \rho_{\psi} \implies \sum_n |a_n^\psi|^2 = 1,
\end{aligned}
\end{equation}
where we employed $\rho_\psi = \rho_\psi^\dagger$ and $A_n^\psi \rho_{\psi} = \rho_{\psi} (A_n^\psi)^\dagger = 0$. 
From this the trace-preservation condition implies
\begin{equation}
\begin{aligned}
\mathds{1} & = \sum_n \left(K_{n}^{\psi}\right)^\dagger K_n^\psi \\
& =\rho_{\psi} + \sum_n \left[ (a_n^\psi)^{*} \rho_{\psi} A_n^\psi + a_n^\psi \left(A_n^\psi \right)^\dagger \rho_{\psi} + \left(A_n^\psi \right)^\dagger A_n^\psi \right].
\end{aligned}
\label{eq:condition}
\end{equation}
For any pure state, one can always write the decomposition $|\phi\rangle = \zeta |\psi\rangle + \zeta_\perp |\psi_\perp\rangle$, where $\langle \psi|\psi_\perp\rangle = 0$ and $|\zeta|^2 + |\zeta_\perp|^2 = 1$, so that $\Sigma_{\psi}(\phi) = D\langle \psi|\phi\rangle\langle \phi|\psi\rangle = D|\zeta|^2$. After the state is acted upon by the free channel, we obtain
\begin{multline}
\Lambda^{\psi}(\phi) = |\zeta|^2 \rho_{\psi} + |\zeta_\perp|^2 \sum_n A_n^{\psi} |\psi_\perp\rangle\langle \psi_\perp| (A_n^{\psi})^\dagger \\
+ \zeta \zeta^{*}_\perp \sum_n a_n^{\psi} |\psi\rangle\langle \psi_\perp| (A_n^{\psi})^\dagger + \zeta^* \zeta_\perp \sum_n (a_n^\psi)^{*} A_n^{\psi} |\psi_\perp\rangle\langle \psi|.
\end{multline}
Consequently,
\begin{equation}
\begin{split}
\Sigma_{\psi}(\Lambda^{\psi}(\phi)) & = D\langle \psi|\Lambda^{\psi}(\phi)|\psi\rangle \\
& = D|\zeta|^2 + |\zeta_\perp|^2 \sum_n |\langle \psi|A_n^{\psi}|\psi_\perp\rangle|^2 \\
& \quad\quad + \left[ \zeta \zeta^{*}_\perp \sum_n a_n^{\psi} \langle \psi_\perp| (A_n^{\psi})^\dagger |\psi\rangle + \text{c.c.} \right],
\end{split}
\label{eq:general_metric}
\end{equation}
where \text{c.c.} stands for \text{complex conjugate}. 

We depart from the original proof by dropping the assumption that $\Sigma_\psi$ corresponds to the grand sum (as for the texture).
Our goal is to show that the terms in the bracket above are zero. To do this we project $\bra{\psi_\perp}$ from the left and $\ket{\psi}$ from the right in Eq.~\eqref{eq:condition}, resulting in
\begin{multline}
0 = \langle\psi_\perp\vert \rho_{\psi} \vert\psi\rangle + \sum_n \bigg[ (a_n^\psi)^{*} \braket{\psi_\perp|\rho_{\psi} A_n^\psi|\psi} \\
+ a_n^\psi \braket{\psi_\perp| \left(A_n^\psi \right)^\dagger \rho_{\psi}|\psi} +\braket{\psi_\perp|\left(A_n^\psi \right)^\dagger A_n^\psi |\psi} \bigg]
\end{multline}
The first term goes to zero through simple orthogonality and the second and fourth terms from $A_{n}^\psi \vert\psi \rangle=0$ as defined above, leaving 
\begin{equation}
\sum_n a_n^\psi \braket{\psi_\perp|\left(A_n^\psi \right)^\dagger |\psi} = 0
\end{equation}
where we used $\rho_{\psi}=\vert\psi\rangle\langle\psi\vert$.
Hence this and its complex conjugate are zero and the bracketed terms in Eq.~\eqref{eq:general_metric} vanish. 
Consequently,
\begin{equation}
\Sigma_{\psi}(\Lambda^{\psi}(\phi)) = \Sigma_{\psi}(\phi) + |\zeta_\perp|^2 \sum_n |\langle \psi|A_n^\psi|\psi_\perp\rangle|^2 \geq \Sigma_{\psi}(\phi).
\end{equation}
Therefore, $\Sigma_{\psi}$ is nondecreasing under arbitrary free operations $\Lambda^\psi$ for pure states. The generalized rugosity $\mathcal{R}_{\psi}$ [\cref{eq:Rugosity}] therefore does not increase under free operations for pure states.
Given that $\Sigma_{\psi}$ is a linear operation, our result that nonincreasing $\cR_\psi$ can be trivially extended to mixed states as
\begin{equation}
\mathcal{R}_{\psi}(\Lambda^{\psi}(\rho)) \leq \mathcal{R}_{\psi}(\rho)
\end{equation}
for arbitrary free maps and density matrices $\rho$, establishing monotonicity. 
Taken together with the nonnegativity of $\mathcal{R}_{\psi}$, meaning $\mathcal{R}_{\psi}(\rho)\geq0$ (and $\mathcal{R}_{\psi}(\rho)=0 \iff \rho=\vert\psi\rangle\langle\psi\vert$), we conclude that it is a resource measure of generalized texture.

\section{Convex-Roof Derivation for Coherence}
\label{app:coherence_derivation}

Here we extend the derivation in Sec.~\ref{sec:basisDepCoh} to arbitrary mixed single-qubit states $\rho$ by minimizing over all pure state decompositions:
\begin{widetext}
\begin{equation}
    \mathcal{R}_{\mathcal{F}}(\rho)=\min_{p,\hat{q},\hat{s}}\left[-p\ln\left(\frac{1+\vert q_{z}\vert}{2}\right)-(1-p)\ln\left(\frac{1+\vert s_{z}\vert}{2}\right)\right]\ge-\ln\left[\frac{p}{2}(1+\vert q_{z}\vert)+\frac{1-p}{2}(1+\vert s_{z}\vert)\right].
\end{equation}
\end{widetext}
Saturation again occurs when the arguments are equal, leading to $\vert q_{z}\vert =\vert s_{z}\vert$. In other words, the two pure states in the decomposition must have the same value of pure state rugosity.
Similar to the imaginarity case we have two solutions, though only explicitly explore one of them below ($q_{z}=-s_{z}$) as the other ($r_{z}=q_{z}=s_{z}$) does not minimize the rugosity. 

To find an explicit construction we rewrite the purity constraints as 
$s_{\perp}^{2}+s_{z}^{2}=q_{\perp}^{2}+q_{z}^{2}=1$, and set the perpendicular components to trivially recover $r_{\perp}$ as $s_{\perp}=q_{\perp}=r_{\perp}$ so that $p$ only mediates the $z$ components. 
Note that this satisfies the Jensen's inequality constraint since $\vert q_{z}\vert=\vert s_{z}\vert=\sqrt{1-r_{\perp}^{2}}$.
We now use the additional constraint $q_{z}=-s_{z}=\sqrt{1-r_{\perp}^{2}}$, which is mediated by $p$ as $r_{z}=pq_{z}+(1-p)s_{z}$ or rewritten for this branch as $r_{z}=(2p-1)q_{z}=(2p-1)\sqrt{1-r_{\perp}^{2}}$. Since for all valid mixed states $\vert r_{z}\vert\leq \sqrt{1-r_{\perp}^{2}}$ this construction is always possible. 
The convex-roof extension is therefore given by (reprinted from the main text for clarity)
\begin{equation}
    \mathcal{R}_{\mathcal{F}}(\rho)=-\ln\left(\frac{1+\sqrt{1-r_{\perp}^{2}}}{2}\right).
\end{equation}

\section{Violation of strong monotonicity}
\label{app:strong_monotonicity}

Here we present a set of $D$-dimensional pure states that, when used with a fixed point map $\Lambda$,  violate strong monotonicity for the measure $\mathcal{R}_{\mathcal{F}_o}$. While we consider the case in which our free set $\mathcal{F}_o$ contains a single state ($\mathcal{F}_o =\{\ket{1}\}$), this example can be extended to resources with free sets of arbitrary dimension.

We must reiterate that the \textit{weak} monotonicity of a measure $M(\rho)$ holds if its value cannot increase under free operations $\Lambda$, i.e. $M(\rho) \geq M(\Lambda(\rho))$ if $M$ is weakly monotonic.  
On the other hand, $M$ is \textit{strongly} monotonic if its average value over outcomes $\sigma_j = \frac{K_j\rho K_j^\dagger}{\text{Tr}\{K_j \rho K_j^\dagger\}}$ of a channel $\Lambda(\rho) = \sum_jK_j\rho K_j^\dagger$ is less than $M(\rho)$ itself: $\sum_j p_j M(\sigma_j) \leq M(\rho)$. Physically, the weights $p_j = \text{Tr}\{K_j \rho K_j^\dagger\}$ represent the probabilities of outcomes $\sigma_j$ occurring.

Consider the pure initial state 
\begin{equation}
    \vert \tau\rangle=\sqrt{a}\vert 1\rangle +\sqrt{\frac{1-a}{D-1}}\sum_{k=2}^{D}\vert k\rangle
\end{equation}
where the $\vert k\rangle$ states form an orthonormal basis and $1/2\le a\le 1$.
This has a rugosity of $\mathcal{R}_{\cF_o}(\tau)=-\ln{a}$.

We now consider two Kraus operators, both diagonal, selected specifically so that the first one is a filter capable of reweighting this state, with some probability, to one of higher rugosity:
\begin{equation}
K_{1}=
\begin{pmatrix}
    \sqrt{\frac{1-a}{a(D-1)}} & 0 \\
    0 & \mathds{1}_{D-1} \\
\end{pmatrix}
\end{equation}
with $K_{2}=\sqrt{\mathds{1}_{D}-K_{1}^{\dagger}K_{1}}$.

The two (unnormalized) output states are given by
\begin{equation}
    \vert T_{1}'\rangle=K_{1}\vert\tau\rangle=\sqrt{\frac{1-a}{D-1}}\sum_{k=1}^{D}\vert k\rangle
\end{equation}
and $\vert T_{2}'\rangle=K_{2}\vert\tau\rangle\propto\vert 1\rangle$.
The probability of the outcome given by $\vert T_{1}'\rangle$ is  
\begin{equation}
    p_{1}=\langle T_{1}'\vert T_{1}'\rangle=\frac{D(1-a)}{D-1}
\end{equation}
and $p_{2}=1-p_{1}$,
leading to the normalized output states
\begin{equation}
    \vert T_{1}\rangle=\frac{1}{\sqrt{D}}\sum_{k=1}^{D}\vert k\rangle,\hspace{3mm} \vert T_{2}\rangle=\vert 1\rangle
\end{equation}
with rugosities $\mathcal{R}_{\cF_o}(T_{1})=-\ln\frac{1}{D}=\ln{D}$ and $\mathcal{R}_{\cF_o}(T_{2})=0$ (since $\ket{T_2}\in\cF_o$). 
The strong monotonicity inequality then becomes
\begin{equation}
\begin{aligned}
    \mathcal{R}_{\cF_o}(\tau)\ge p_{1}\mathcal{R}_{\cF_o}(T_{1})+p_{2}\mathcal{R}_{\cF_o}(T_{2})\\
    -\ln{a}\ge \frac{D(1-a)}{D-1}\ln{D}\\
\end{aligned}
\end{equation}
This will be violated whenever
\begin{equation}
    aD^{\frac{D(1-a)}{D-1}}>1.
\end{equation}
It is easy to find violations, for example, the family of solutions given by  $a(D)=1-\frac{1}{2D}$ violates for all the dimensions we checked (e.g., $D=2,3, 10$). Specifically:
\begin{equation}
    \begin{aligned}
        a(2)&=\frac{3}{4}\ \rightarrow\ \frac{3}{4}2^{\frac{1}{2}}\approx 1.06>1\\
        a(3)&=\frac{5}{6}\ \rightarrow\ \frac{5}{6}3^{\frac{1}{4}}\approx 1.10>1\\
        a(10)&=\frac{19}{20} \rightarrow\ \frac{19}{20}10^{\frac{1}{18}}\approx 1.08>1.
    \end{aligned}
\end{equation}

\section{Numerical Validation of Convex-Roof Measures}
\label{app:numerical}
In Sec.~\ref{sec:relationship_with_measures}, we generalize the original proposed texture measure to allow multiple states of zero resource [Eq.~\eqref{eq:rugosityPure}] and rugosity definition of mixed states based on the convex-roof extension [Eq.~\eqref{eq:convex_roof_exact}]. The computational demands for evaluating Eq.~\eqref{eq:convex_roof_exact} depend both on the choice of free states $\mathcal{F}$ and on the dimension and rank of our states $\rho$. 
A more detailed discussion of the convex-roof extension and its calculation can be found within \cite{uhlmann2010roofs,zhu2025unified,PhysRevA.111.032435,leditzky2017useful}. In what follows, we numerically evaluate the convex-roof extension using nonconvex means (differential evolution \cite{PhysRevA.111.032435,storn1997differential}) to demonstrate the weakly monotonic properties of our measure $\mathcal{R}_{\mathcal{F}_o}$ with the application of random free operations for special cases of fixed-point resource theories. 

\subsection{Generating Random Free Operations, Evaluating the Convex-Roof Extension} 
We shall consider a single-particle, four-level system ($D=4$) to study the weak monotonicity properties of the measure $\mathcal{R}_{\mathcal{F}_o}(\rho)$ after repeated free operations $\Lambda$ for free states $\psi_i$ of dimension $\dim\{\mathcal{F}_o \}\leq 3$. 
One could also consider the special case of genuine coherence $\dim\{\psi_i\}= D = 4$, but we have left this study to \cite{de2016genuine}. 
In this example, our free states correspond to computational bases, i.e. $\ket{\psi_i} \in \{\ket{i}\}_{i\in[D]}$, but one can also consider free states in an arbitrary basis. 
For simplicity, we use single qudit states, but our measure can be applied to systems of arbitrary dimension and number of particles. The interpretations of our results become substantially more complicated for multipartite systems, in which ``laboratory independence'' of free operations depends on a given subsystem (resulting from Theorem 1 in \cite{de2016genuine}).

As discussed in Sec.~\ref{sec:fixed}, the Kraus operators associated with free operations in a fixed-point resource theory can be written in the block-triangular form
\begin{equation}
\label{eq:krausAppB}
    K_n  \;=\;
\begin{pmatrix}
  \text{diag}(\alpha_{1,n},\alpha_{2,n},...,\alpha_{m,n}) & T_{n} \\
  \mathbf{0}   & L_n
\end{pmatrix},
\end{equation}
where $m=\text{dim}\{\mathcal{H}_{\mathcal{F}_o}\}$.
To enforce trace-preservation, these Kraus operators $\{K_n\}$ must additionally satisfy
\begin{equation}
\label{eq:tracePres}
\sum_n K_n^\dagger K_n=\mathds{1}.
\end{equation}
Generating random matrices that satisfy \emph{both} Eqs.~\eqref{eq:krausAppB} and \eqref{eq:tracePres} is nontrivial; we implement a solution combining an algebraic trick with the Schur complement~\cite{boyd2004convex}. 

Consider first the operators
\begin{equation}
\label{eq:krausAppBtilde}
    \widetilde{K}_n  \;=\;
\begin{pmatrix}
  \text{diag}(\tilde{\alpha}_{1,n},\tilde{\alpha}_{2,n},...,\tilde{\alpha}_{m,n}) & \widetilde{T}_{n} \\
  \mathbf{0}   & \widetilde{L}_n
\end{pmatrix}
\end{equation}
with $\widetilde{T}_{n}\in\mathbb{C}^{m\times(D-m)}$ and $\widetilde{L}_n\in\mathbb{C}^{(D-m)\times(D-m)}$. Sampling each nonzero element in \cref{eq:krausAppBtilde} from a standard complex normal distribution produces a set of block-triangular matrices that resolve to some Hermitian matrix that we can also express in a block form:
\begin{equation}
    \sum_n \widetilde{K}_n^\dagger \widetilde{K}_n = 
    \begin{pmatrix}
    M_{00} & M_{01} \\
    M_{01}^\dagger & M_{11}
    \end{pmatrix}.
\end{equation}
We next define
\begin{equation}
    X = 
    \begin{pmatrix}
    M_{00}^{-1/2} & -M_{00}M_{01}S^{-1/2} \\
    \mathbf{0} & S^{-1/2}
    \end{pmatrix},
\end{equation}
where $S=M_{11}-M_{01}^\dagger M_{00}^{-1}M_{01}$ represents the Schur complement of block $M_{00}$. Then, the the Kraus operators defined by
\begin{equation}
\label{eq:KrausNorm}
    K_n = \widetilde{K}_n X
\end{equation}
are guaranteed to be both block-triangular [Eq.~\eqref{eq:krausAppB}] and trace-preserving [Eq.~\eqref{eq:tracePres}]. 

For our simulations, we apply randomly generated free operations [Eqs.~(\ref{eq:krausAppBtilde}--\ref{eq:KrausNorm})] recursively such that output state after the $i^\text{th}$ free operation is 
\begin{equation}
    \rho^{(i)} = \Lambda^{(i)}(\rho^{(i-1)}),
    \label{eq:recrho}
\end{equation}
and $\rho^{(0)} = \ket{\phi}\!\!\bra{\phi} = \ket{4}\bra{4}$ (we have chosen this initial state since it is maximally resourceful in \textit{all} cases presented in this section).
The recursive definition of Eq. \eqref{eq:recrho} has been chosen specifically to illustrate the deleterious effect of consecutive free operations, noting that the composition of free operations is also a free operation.
Following the Kraus definitions in Eq.~\eqref{eq:KrausNorm}, the free operations can be defined as $\Lambda^{(i)}(\rho) = \sum_{j=1}^R K_{j}^{(i)}\rho (K_{j}^{(i)})^\dagger$, where the Kraus rank is $R=3$ for the purpose of demonstration. For each iteration, $\rho^{(i)}$, we evaluate the measure $M(\rho^{(i)})=\mathcal{R}_{\mathcal{F}_o}(\rho^{(i)})$ defined in Eq.~\eqref{eq:convex_roof_exact}. 

While the pure-state measures $\mathcal{R}_{\mathcal{F}_o}(\phi)$ [Eq.~\eqref{eq:rugosityPure}] are directly computable, the convex-roof extension proves a more complicated task because of the nonconvexity of finding the optimal decomposition of pure states $\ket{\phi_j}$ \cite{PhysRevA.111.032435}. A naïve approach would involve a global search across all possible decompositions of a state $\rho$, requiring both a parameterization of pure states in the decomposition and their respective weights. We note that although a state $\rho$ has infinitely many decompositions, one can place an upper bound on the number states in a decomposition with the help of Carathéodory's theorem~\cite{streltsov2010linking}. Such a search would also require one to impose a soft penalty to ensure that a derived decomposition indeed corresponds to the state $\rho$, making the search intractable for increasing system sizes due to the quadratic scaling of states required in a decomposition, predicted by Carathéodory's theorem.

Instead, we use a result from the Schrödinger-HJW theorem \cite{kirkpatrick2006schrodinger}, where the decompositions of a state $\rho$ can be parameterized by a set of $k\times r$ matrices $U\in \mathds{C}^{k\times r}$~\cite{PhysRevA.111.032435, PhysRevA.80.042301}.  The value $k$ is the number of states in a given decomposition (bounded by Carathéodory's theorem such that $k\leq D^2$) and $r$ denotes the rank of $\rho$. Importantly, $U$ must have the property $U^\dagger U = \mathds{1}_{r}$. Later in this section, we will show that searching over matrices $U$ is equivalent to searching over the set of $k\times k$ unitary matrices $Q\in U(k)$, which reduces the convex-roof extension problem to a global search over the surface of a manifold, without the need for soft penalties. 
A summary of our workflow can be found in Algorithms \ref{alg:schhrodinger_hjw} and \ref{alg:convex_roof}. 

\begin{algorithm}[H]
\caption{Schrödinger-HJW Decomposition}
\label{alg:hjw}
\begin{algorithmic}[1]
\State \textbf{Input:} Mixed state $\rho$, semi-unitary matrix $U\in \mathds{C}^{k\times r}$ 
\State $\{\lambda_i,\ket{v_i}\} \gets \text{eig}(\rho)$ \Comment{Diagonalize $\rho$}
\State $\ket{\zeta_i} \gets \sum_j U_{ij}\sqrt{\lambda_j}\ket{v_j}$
\State $p_i \gets \braket{\zeta_i|\zeta_i}$
\State $\ket{\phi_i} \gets \frac{1}{\sqrt{p_i}}\ket{\zeta_i}$

\State \textbf{Output:} $\left\{p_i,\ket{\phi_i}\right\}$
\end{algorithmic}
\label{alg:schhrodinger_hjw}
\end{algorithm}

\begin{algorithm}[H]
\caption{Convex-Roof Extension \(\mathcal{R}_{\mathcal{F}_o}(\rho)\)}
\label{alg:convex_roof}
\begin{algorithmic}[1]
\State \textbf{Input:} Mixed state \(\rho\), free states \(\{\psi_i\}\), decomposition size \(k\). The rank of $\rho$ is represented as $r$.
\Function{$\mathrm{Obj}$}{$\mathbf{x}$} \Comment{Objective function\\Parameter $\mathbf{x}$ represents a vectorization of a $k\times r$ matrix $X$}

\State $X \gets x.\text{reshape}\left(k,r\right)$
\State $Q,R \gets \text{qr}(X)$ \Comment{QR decomposition}
\State $U \gets Q_{[:,r]}$
\State $\{p_i,\ket{\phi_i}\} \gets \text{hjw\_decomp}(\rho,U)$
\State \textbf{return} $\sum_{i} p_i\mathcal{R}_{\mathcal{F}_o}(\phi_i)$
\EndFunction
\State Use differential evolution to minimize \(\mathrm{Obj}(\mathbf{x})\)
\end{algorithmic}
\end{algorithm}

The most important component of our pipeline is the objective function in Algorithm~\ref{alg:convex_roof}, which accepts a vector $\mathbf{x}\in\mathbb{C}^{kr\times 1}$ corresponding to a ``flattened'' version of the $k\times r$ matrix $X$---required because the SciPy implementation of differential evolution only accepts a one-dimensional array as input to a user's objective function. We can obtain a $k\times k$ unitary matrix $Q$ from $X$ by QR decomposition: 
\begin{equation}
    X = QR,
\end{equation}
where $R$ is a $k\times r$ right-triangular matrix. We convert $Q$ to a $k\times r$ semi-unitary matrix $U$ by discarding its last $k-r$ columns: $U = Q_{[:,r]}$. Following this, we can straightforwardly define a decomposition of $\rho$ according to 
\begin{equation}
    \rho = \sum_{i=1}^k p_i\ket{\phi_i}\bra{\phi_i},
\end{equation}
where $p_i = \braket{\zeta_i|\zeta_i}$, $\ket{\phi_i} = \frac{1}{\sqrt{p_i}}\ket{\zeta_i}$ and crucially, 
\begin{equation}
    \ket{\zeta_i} = \sum_{j=1}^r U_{ij}\sqrt{\lambda_{j}}\ket{v_j},
\end{equation}
where $\lambda_j$ and $\ket{v_j}$ are the respective eigenvalues and eigenstates of $\rho$. Using the HJW-decomposition, the objective function returns the evaluated convex-roof extension for a choice of semi-unitary matrix $U$. We minimize our objective function using differential evolution, as implemented in the SciPy library, with a maximum number of $500$ iterations and a population size of 40.

\begin{figure*}[tb!]
    % \centering
    \begin{tikzpicture}[scale=1]
        \begin{axis}[
            xlabel={No. of Successive Free Operations},
            ylabel={Resource Measure},
            xmin=0, xmax=5,
            ymin=0, ymax=2.5,
            xtick={0,1,2,3,4},
            domain = {0:4},
            ytick={0.5,1,1.5,2,2.5},
            legend pos=north east,
            ymajorgrids=true,
            xmajorgrids=true,
            grid style=dashed,
            error bars/y dir=both, % turn on error bars
            error bars/y explicit,  % say that error value is given explicitly
            width=0.7\textwidth,
            height=0.4\textwidth
        ]   
        
        \addplot[mark=nomark,color=color1bg, mark size=1pt,dashed, mark options={thin,solid},error bars/.cd, x dir=none, y dir=both, y explicit,
            error bar style={color=color1bg,solid, line width = 1 pt}, error mark options = {line width = 8 pt}] 
            table [x expr = \thisrow{Iteration}, y expr =\thisrow{Measure}, y error plus expr = \thisrow{Measure_High} - \thisrow{Measure}, y error minus expr = \thisrow{Measure} - \thisrow{Measure_Low}, col sep=comma, mark=square] {./data/convex_roof_freedims_processed_d1_20260207_185249.csv};

        \addplot[mark=nomark,color=color3bg, mark size=1pt,dashed, mark options={thin,solid},error bars/.cd, x dir=none, y dir=both, y explicit,
            error bar style={color=color3bg,solid, line width = 1 pt}, error mark options = {line width = 8 pt}] 
            table [x expr = \thisrow{Iteration}, y expr =\thisrow{Measure}, y error plus expr = \thisrow{Measure_High} - \thisrow{Measure}, y error minus expr = \thisrow{Measure} - \thisrow{Measure_Low}, col sep=comma, mark=square] {./data/convex_roof_freedims_processed_d2_20260207_185249.csv};

        \addplot[mark=nomark,color=color5bg, mark size=1pt,dashed, mark options={thin,solid},error bars/.cd, x dir=none, y dir=both, y explicit,
            error bar style={color=color5bg,solid, line width = 1 pt}, error mark options = {line width = 8 pt}] 
            table [x expr = \thisrow{Iteration}, y expr =\thisrow{Measure}, y error plus expr = \thisrow{Measure_High} - \thisrow{Measure}, y error minus expr = \thisrow{Measure} - \thisrow{Measure_Low}, col sep=comma, mark=square] {./data/convex_roof_freedims_processed_d3_20260207_185249.csv};
        \end{axis}
    \end{tikzpicture}

    \begin{tikzpicture}[]
    \begin{axis}[
        hide axis,
        width = 0.9\textwidth,
        height = 0.1\textwidth,
        xmin=0, xmax=1,
        ymin=0, ymax=1,
        legend columns=3,
        legend style={
            at={(0.5,-0.25)},
            anchor=north,
            column sep=2em,
            draw=none
        },
        every legend image/.style={
            error bar legend,
        },
        every node/.append style={
            draw, rounded corners, inner sep=2pt
        }
    ]

    \addlegendimage{error bar legend={color1bg}}
    \addlegendentry{$\mathcal{R}_{\mathcal{F}_o}(\rho)$, $\dim \{\mathcal{F}_o\}=1$}

    \addlegendimage{error bar legend={color3bg}}
    \addlegendentry{$\mathcal{R}_{\mathcal{F}_o}(\rho)$, $\dim \{\mathcal{F}_o\}=2$}
    \addlegendimage{error bar legend={color5bg}}
    \addlegendentry{$\mathcal{R}_{\mathcal{F}_o}(\rho)$, $\dim \{\mathcal{F}_o\}=3$}

    \end{axis}
\end{tikzpicture}

    \caption{
    \justifying
    Convex-roof extension $\mathcal{R}_{\mathcal{F}_o}(\rho^{(i)})$ evaluated for initial state $\rho^{(0)}=\ket{4}\bra{4}$ after $i$ rounds of successive, randomly generated free operations $\Lambda^{(i)}$. $\mathcal{R}_{\mathcal{F}_o}(\rho^{(i)})$ decays monotonically with increasing number of operations for all dimensions $\dim\{\mathcal{F}_o\} \in\{ 1,2,3\}$ corresponding to blue, green, and purple colors respectively. Datapoints connected by dashed lines correspond to the statistical mode of the prediction of our differential evolution algorithm after 30 repeated trials. Error bars correspond to quantiles within the range [0.05, 0.95]. 
    The numerical observation $\mathcal{R}_{\mathcal{F}_o}(\rho^{(i)}) \leq \mathcal{R}_{\mathcal{F}_o}(\rho^{(i-1)})$ is consistent with weak monotonicity.}
    \label{fig:rugosity_data}
\end{figure*}
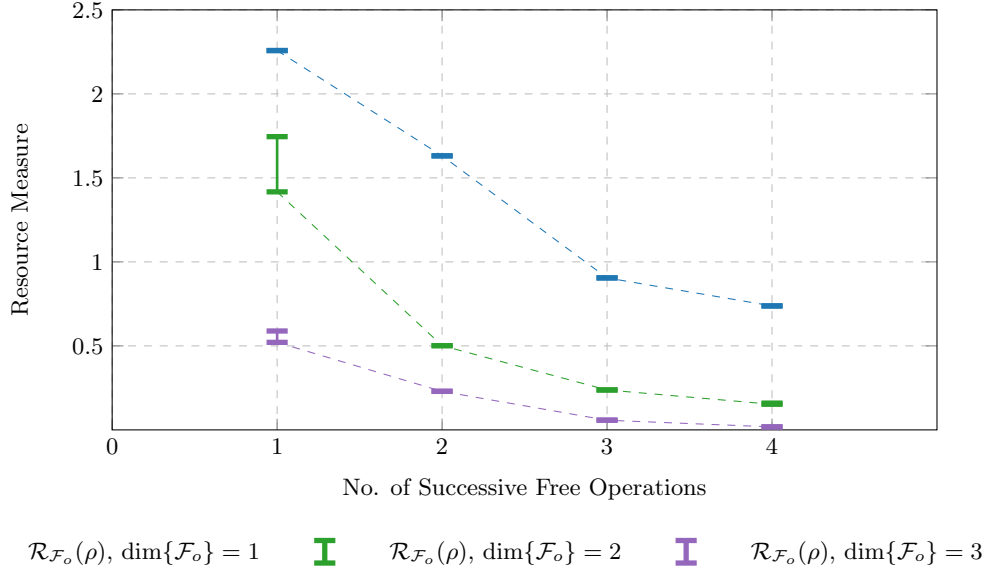

Figure~\ref{fig:rugosity_data} shows the numerical data of this study after performing up to four consecutive rounds of random free operations $\Lambda^{(i)}$, effectively performing a random walk. Data points connected by dashed lines correspond to the statistical mode of the predictions of our stochastic algorithm after 30 repeated trials. The error bars denote the 5th--95th quantile range, highlighting the highly skewed nature of the distributions in some instances.
In all cases, $\mathcal{R}_{\mathcal{F}_o}(\rho^{(i)})$ decreases \textit{monotonically} with the application of additional free operations when, which is a signature of $\mathcal{R}_{\mathcal{F}_o}$ possessing weak monotonicity. 

\subsection{Example Data} 
As part of our demonstration of evaluating the convex-roof extension, we present some of the data used in the $\dim\{\cF_o\}=1$ case of \cref{fig:rugosity_data}. The Kraus operators used in the first operation $\Lambda^{(1)}$ of the first of the 30 trials are listed in Eqs.~(\ref{eq:k0_example}--\ref{eq:k2_example}). We note that each ``round" of free operations consists of an entirely new randomly generated operation $\Lambda^{(i)}$, meaning that the Kraus operators presented in this section are only representative of the states and operations considered in the first datapoint of the \textit{blue} plot in \cref{fig:rugosity_data}.
\begin{widetext}

\begin{equation}
K_{1}^{(1)} =
\begin{array}{c@{\hspace{5pt}}c}
 & \begin{array}{*{4}{c}}
   \bra{1} & \bra{2} & \bra{3} & \bra{4}
   \end{array} \\[5pt]
\begin{array}{c}
\ket{1} \\ \ket{2} \\ \ket{3} \\ \ket{4}
\end{array}
&
\left(\begin{array}{*{4}{c}}
0.506-0.278\imag & -0.005-0.075\imag & 0.037+0.167\imag & 0.046-0.173\imag \\
0 & -0.292+0.095\imag & -0.143-0.036\imag & 0.061-0.414\imag \\
0 & 0.446-0.148\imag & -0.182-0.248\imag & 0.071-0.115\imag \\
0 & -0.042-0.028\imag & -0.089-0.356\imag & 0.075-0.115\imag
\end{array}\right)
\end{array}
\label{eq:k0_example}
\end{equation}

\begin{equation}
K_{2}^{(1)} =
\begin{array}{c@{\hspace{5pt}}c}
 & \begin{array}{*{4}{c}}
   \bra{1} & \bra{2} & \bra{3} & \bra{4}
   \end{array} \\[5pt]
\begin{array}{c}
\ket{1} \\ \ket{2} \\ \ket{3} \\ \ket{4}
\end{array}
&
\left(\begin{array}{*{4}{c}}
0.484-0.315\imag & -0.106+0.260\imag & 0.259-0.145\imag & 0.090-0.042\imag \\
0 & 0.142-0.028\imag & 0.047+0.108\imag & 0.458+0.266\imag \\
0 & -0.260+0.345\imag & -0.284+0.057\imag & 0.145+0.445\imag \\
0 & 0.235-0.151\imag & 0.135+0.377\imag & -0.221+0.143\imag
\end{array}\right)
\end{array}
\label{eq:k1_example}
\end{equation}

\begin{equation}
K_{3}^{(1)} =
\begin{array}{c@{\hspace{5pt}}c}
 & \begin{array}{*{4}{c}}
   \bra{1} & \bra{2} & \bra{3} & \bra{4}
   \end{array} \\[5pt]
\begin{array}{c}
\ket{1} \\ \ket{2} \\ \ket{3} \\ \ket{4}
\end{array}
&
\left(\begin{array}{*{4}{c}}
0.520-0.250\imag & 0.140-0.170\imag & -0.303-0.058\imag & -0.150+0.201\imag \\
0 & -0.126-0.196\imag & -0.263+0.330\imag & 0.197-0.153\imag \\
0 & 0.378+0.159\imag & -0.024+0.156\imag & 0.044-0.041\imag \\
0 & 0.204-0.009\imag & 0.234+0.145\imag & -0.212+0.058\imag
\end{array}\right)
\end{array}
\label{eq:k2_example}
\end{equation}
\end{widetext}

The state decomposition of $\rho^{(1)}$ deemed optimal by our global search can be found in Table \ref{tab:state_decomposition}, where a state obtained in round $i$ takes the form: $\rho^{(i)} = \sum_{j=1}^kp_j^{(i)}\ket{\phi_j^{(i)}}\bra{\phi_j^{(i)}}$, with $\ket{\phi_j^{(i)}} = \alpha_j^{(i)}\ket{1} + \beta_j^{(i)}\ket{2} + \gamma_j^{(i)}\ket{3} + \delta_j^{(i)}\ket{4}$. Although each datapoint in Fig. \ref{fig:rugosity_data} is representative of the results across 30 trials, we choose state decompositions that lie in the statistical mode of the predicted measure $\mathcal{R}_{\mathcal{F}_o}$. 

\begin{table*}[tb!]
\centering
\begin{tabular}{p{4em}p{5em}p{8em}p{8em}p{8em}p{8em}}
\hline \hline
 & $p_{j}^{(1)}$ & $\alpha_{j}^{(1)}$ & $\beta_{j}^{(1)}$ & $\gamma_{j}^{(1)}$ & $\delta_{j}^{(1)}$ \\ \hline
$\phi_1^{(1)}$ & 0.149 & $-0.266+0.183\imag$ & $-0.361-0.615\imag$ & $-0.033-0.610\imag$ & $0.090-0.077\imag$ \\
$\phi_2^{(1)}$ & 0.021 & $0.198+0.256\imag$ & $-0.016-0.761\imag$ & $0.182-0.025\imag$ & $-0.158+0.507\imag$ \\
$\phi_3^{(1)}$ & 0.022 & $0.281-0.161\imag$ & $0.110+0.539\imag$ & $-0.156-0.176\imag$ & $0.648-0.343\imag$ \\
$\phi_4^{(1)}$ & 0.091 & $0.263+0.188\imag$ & $0.684-0.360\imag$ & $0.365+0.078\imag$ & $0.046+0.395\imag$ \\
$\phi_5^{(1)}$ & 0.065 & $0.273+0.173\imag$ & $0.427+0.602\imag$ & $-0.220+0.501\imag$ & $-0.224+0.028\imag$ \\
$\phi_6^{(1)}$ & 0.069 & $-0.181+0.268\imag$ & $-0.610+0.516\imag$ & $-0.428-0.222\imag$ & $0.042-0.153\imag$ \\
$\phi_7^{(1)}$ & 0.094 & $0.088-0.311\imag$ & $0.147-0.705\imag$ & $0.441-0.245\imag$ & $0.336+0.097\imag$ \\
$\phi_8^{(1)}$ & 0.072 & $0.322+0.029\imag$ & $0.546+0.293\imag$ & $0.337+0.435\imag$ & $0.053+0.454\imag$ \\
$\phi_9^{(1)}$ & 0.109 & $-0.211-0.245\imag$ & $0.453+0.530\imag$ & $0.166+0.509\imag$ & $-0.337-0.095\imag$ \\
$\phi_{10}^{(1)}$ & 0.040 & $-0.243+0.214\imag$ & $-0.766-0.364\imag$ & $-0.082-0.362\imag$ & $-0.066+0.187\imag$ \\
$\phi_{11}^{(1)}$ & 0.036 & $0.217-0.240\imag$ & $-0.238+0.060\imag$ & $-0.388-0.331\imag$ & $0.488-0.581\imag$ \\
$\phi_{12}^{(1)}$ & 0.059 & $-0.241+0.215\imag$ & $-0.220+0.664\imag$ & $-0.468+0.098\imag$ & $-0.298-0.297\imag$ \\
$\phi_{13}^{(1)}$ & 0.026 & $-0.206+0.249\imag$ & $-0.758-0.344\imag$ & $-0.227-0.170\imag$ & $-0.312+0.157\imag$ \\
$\phi_{14}^{(1)}$ & 0.067 & $0.323-0.015\imag$ & $-0.788-0.300\imag$ & $-0.420-0.068\imag$ & $0.054-0.012\imag$ \\
$\phi_{15}^{(1)}$ & 0.027 & $-0.044-0.320\imag$ & $-0.451+0.664\imag$ & $-0.303+0.210\imag$ & $0.032-0.339\imag$ \\
$\phi_{16}^{(1)}$ & 0.053 & $0.312+0.084\imag$ & $0.461-0.350\imag$ & $0.465-0.080\imag$ & $0.353+0.461\imag$ \\
\hline
\end{tabular}
\caption{\justifying A sample of the decomposition deemed optimal by the differential evolution algorithm used for expressing $\rho^{(1)}$ in one trial of the $\dim\{\cF_0\}=1$ case in the \textit{blue} plot of \cref{fig:rugosity_data}. We have intentionally used a total of $k=16$ states in the decomposition to adhere to the prediction by Carathéodory's theorem ($k\leq D^2$). In this example, $\mathcal{R}_{\mathcal{F}_o}(\rho^{(1)}) \approx 2.258$.}
\label{tab:state_decomposition}
\end{table*}

\end{document}